\documentclass[twocolumn]{article}
\usepackage{epsfig}
\begin{document}
\title{
{\Large\bf  Theory of Myelin Coiling}
}                
\author{\normalsize  Jung-Ren Huang\footnote{email: jhuang2@uchicago.edu}
                       \\
{\sl \small James Franck Institute and Department of Physics,
   University of Chicago} \\
{\sl \small 5640 S. Ellis Avenue, Chicago, Illinois 60637}
}
{\small    \date{\today}  }
\maketitle
\begin{minipage}[b]{6.8in}
\begin{center}
\begin{quotation}
{\bf\noindent Abstract.} 
A new model is proposed to explain coiling of myelins 
composed of fluid bilayers. This model allows 
the constituent bilayer cylinders of a myelin to be 
non-coaxial and the bilayer lateral tension to vary 
from bilayer to bilayer.  The calculations show that
a myelin would bend or coil to lower its free energy 
when the bilayer lateral tension is sufficiently large.
From a mechanical point of view, the proposed coiling
mechanism is analogous to the classical Euler buckling 
 of a thin elastic rod under axial compression. 
 The analysis of a simple two-bilayer case suggests
 that a bilayer lateral tension of about 1 dyne/cm can
 easily induce coiling of myelins  of typical lipid bilayers. 
This model signifies the importance of bilayer lateral
tension in determining the morphology of myelinic structures.
\vspace{0.1in}

\end{quotation}
\end{center}
\end{minipage}
\section{Introduction}
\begin{figure}
\begin{center}
\vspace{2.4in}
\resizebox{0.45\columnwidth}{!}{
    \includegraphics{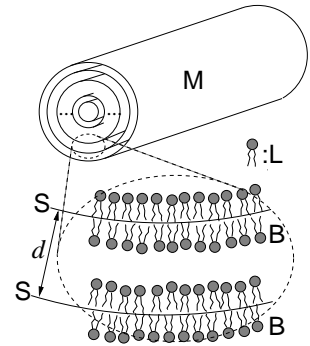}
    }
\end{center}
\caption{    Sketch of myelin structure.
    Myelins  (M) are multilamellar tubes composed of nested
    bilayer cylinders. 
    Oval inset shows arrangements of surfactant 
    molecules (L) in bilayers (B) \cite{Mouritsen-2005,Israelachvili-1993}, 
    where the mid-surface (S) of a bilayer separates
    the two opposing monolayers.
    The symbol $d$ denotes the inter-bilayer spacing.
    In the upper drawing (M), only the bilayer
    mid-surfaces are shown.
    Typical value of bilayer thickness or inter-bilayer 
    spacing is about several 
    nanometers \cite{Mouritsen-2005,Israelachvili-1993} and 
    that of myelin diameter ranges from a few micrometers 
    to about 50 micrometers \cite{Benton-1986,Sakurai-1990,Buchanan-2000,Haran-2002,Zou-2006}.  
\label{fig:myelin_structure}
}
\end{figure}

Surfactant molecules such as  polar lipids
self-assemble into fluid bilayers when dissolved in water 
at sufficiently high temperatures \cite{Mouritsen-2005,Israelachvili-1993}.
If the surfactant concentration is large enough,  
bilayers stack to form multilayer structures called lamellae 
or multilamellae ($L_\alpha$ phase) \cite{Laughlin-1996}.
In some experiments \cite{Harbich-1984,Benton-1986,Sakurai-1990,Buchanan-2000,Haran-2002,Zou-2006}, 
bilayers curve collectively to form nested cylindrical multilamellae known as myelin
figures or simply, myelins (Fig. \ref{fig:myelin_structure}).
Under certain conditions, myelins may bend to 
form coils or double 
helices \cite{Haran-2002,Lin-1982,Sakurai-1985,Mishima-1992,Frette-1999,Dave-2003}.
Because myelins have potential applications such as
controlled drug delivery \cite{Lasic-1993}, detailed 
knowledge of their structure and behavior is desired.
Recently, a geometrical model was proposed to account for 
formation of myelins \cite{Huang-2005}.
This model suggests that if the inter-bilayer repulsion 
is large enough, a flat multilamella is unstable against 
myelin formation, i.e., it would transform into a myelin or 
myelins in order to lower its energy.
In this paper we will focus on coiling of myelins 
composed of fluid bilayers 
and present a theory to explain why they coil.

Several theories of myelin coiling have been proposed.
Lin {et al.} reported that the presence of enough Ca$^{2+}$
ions led to coiling of the myelins  of binary 
mixtures of cardiolipin and phosphatidylcholines \cite{Lin-1982}.
Mishima and his co-workers studied the double helix formation
of the myelins composed of egg-yolk phosphatidylcholine 
bilayers \cite{Mishima-1992}. The authors of these two works
attributed the observed coiling to surface adhesion.
Frette {et al.} investigated coiling of the myelins 
 of phospholipid bilayers 
in the presence of anchored polymers \cite{Frette-1999}.
They suggested that coiling was due to a coupling
between local bilayer curvature and polymer concentration. 
In addition to the above theories, Santangelo and Pincus
proposed a myelin model and showed that myelins are 
unstable to coiling when the bilayer spontaneous curvature 
is nonzero or when the equilibrium 
inter-bilayer spacing is decreased \cite{Santangelo-2002}.
Here we develop a new model. Our model allows the 
bilayer cylinders of a myelin to be non-coaxial and
the bilayer tension to vary from bilayer to 
bilayer \cite{Evans-book,Helfrich-1984}.
By taking account of these two degrees of freedom, 
we will show that a sufficiently large bilayer 
 tension may cause a myelin to coil.
The coiling mechanism proposed here is similar to the 
classical Euler buckling of a thin elastic rod 
under axial load \cite{Love-1944}.

This paper is organized as follows.
In Section \ref{sec:contact} we describe the 
experimental observations that lead to our model.
The proposed coiling mechanism is explained with
a  two-dimensional example 
in Section \ref{sec:bending_2D_tube}.
The definition of our myelin coiling model 
is given in Section \ref{sec:model}.
We  study coiling of a myelin composed of multiple 
bilayers in Section \ref{sec:coiling_N_myelin}.
Coiling of a two-bilayer myelin  is
investigated closely in Section \ref{sec:coiling_2_myelin}.
In Section \ref{sec:Discussion}
we discuss important features and implications of our model.
Finally, Section \ref{sec:conclusions} concludes our work.

%
%
\section{Contact Experiment and Myelin Structure\label{sec:contact}}
In order to provide the motivation for this work,
we will examine closely how myelins are created as well as
their internal structure. We will show that the lateral 
tension of bilayers \cite{Evans-book,Helfrich-1984} should be 
taken into consideration and moreover, the constituent 
bilayer cylinders of a myelin need not be concentric.

\begin{figure}
\begin{center}
\resizebox{0.7\columnwidth}{!}{
    \includegraphics{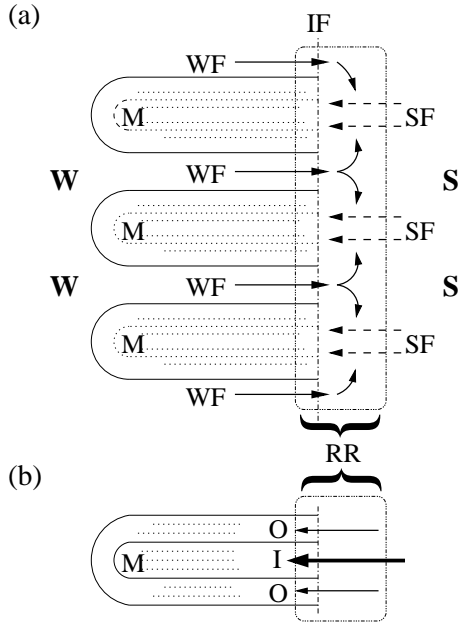}
    }
\end{center}
\caption{
    Sketch of myelin growth and material flux 
    in a contact experiment.
(a): Myelins (M) grow from the interface (IF) of water 
({\bf W}) and bulk surfactant ({\bf S}). 
 The water flux (WF) and surfactant flux (SF) supporting
 myelin  growth enter myelins via the root region (RR).
Solid and dash arrows  indicate the directions of
the water and surfactant fluxes, 
respectively \cite{Buchanan-2000,Haran-2002}. 
(b): A single myelin (M) is naively divided into
a core (I) and a shell (O). 
 As represented by the lengths of the respective arrows,
the material flux densities for these two parts can be 
different. The bilayers of the shell would be in tension
if the material flux density for the core is larger than 
that for the shell.
 \label{fig:contact}
} 
\end{figure}

Myelins are multilamellar tubes 
composed of nested cylindrical fluid bilayers 
(Fig. \ref{fig:myelin_structure}).
They are usually produced in contact 
experiments \cite{Sakurai-1990,Buchanan-2000,Haran-2002}: 
water is brought into contact with
a lump of concentrated surfactant at temperatures high 
enough so that the bilayers are in fluid state. 
Upon contact myelins form at the interface
and grow into water.
The growth of myelins is roughly diffusion-like initially
and slows down significantly after a few 
minutes \cite{Sakurai-1985,Dave-2003,Mishima-1987}.
While myelins are still growing, coiling of single myelins 
or double helix formation of two intertwisting myelins
are often observed \cite{Sakurai-1985}.
Experiments show that as myelins grow, water enters myelins 
mainly via the root region  (Fig. \ref{fig:contact}a)
 \cite{Buchanan-2000,Haran-2002}.
Thus, the influx of  material, surfactant 
and water, that supports myelin growth comes from the 
back of myelins, { i.e.}, the root region.
Because the root region contains  defects of  
sizes smaller than typical myelin 
diameters \cite{Mishima-1987,Sein-1996}, 
the material flux density for a growing myelin is 
expected to be nonuniform\footnote{
The nonuniformity of the material flux density 
depends on the defect
structure in the root region, which would evolve over time 
since defects are not energetically favorable.
}.
Figure \ref{fig:contact}b provides a simple and idealized 
example of nonuniform material influx density. 
In this example the myelin is arbitrarily divided into 
two parts, a core and a shell.
The defect structure in the root region  
 is such that  the material flux density for the 
shell is less than that for the core, 
which implies that the shell is being stretched axially 
and the core compressed also axially.
 Therefore the bilayer tension is expected to be
 {\em nonuniform} and furthermore, the bilayers in the 
 shell should be under tension.
We note that the fluidity of a bilayer leads to
 a uniform tension throughout the bilayer.
In contrast, the notion of ``{\em nonuniform} tension'' used 
in this work always means the tension may vary from bilayer
to bilayer.

Intuitively the bilayer tension in a myelin composed of
fluid bilayers should have some degree of nonuniformity, 
regardless of how the myelin is prepared.
Like an air bubble in water, where the 
interfacial tension is balanced by a pressure difference
across the air-water interface (Laplace's formula),
at least the outermost bilayer of a myelin should be
under tension in order to cancel out the repulsion from the
neighboring bilayer and thus maintain its cylindrical
shape of the body and spherical shape of the end cap(s).
If one or some of the bilayers  are under tension,
the force balance for the myelin body or the end cap(s) 
 implies that the bilayer tension 
may be nonuniform, { i.e.}, it may change from 
bilayer to bilayer.

From the above arguments we expect that the 
bilayer lateral tension  can be variable not only from 
bilayer to bilayer but also from myelin to myelin.
A general myelin model should take 
this degree of freedom into account.
Myelins of uniform bilayer tension are just
special cases of such a model. 
In addition, we will not presume the  bilayer
cylinders of a myelin to be concentric or coaxial
since there is no reason to do so, 
especially for coiled myelins.
Hence, the {\em eccentricities} of the bilayer cylinders
appear in our myelin model naturally.
We will show how these two
variables, the bilayer lateral tension and cylinder 
eccentricity, can give rise to myelin coiling.

%
%
%
\section{Bending of a Two-dimensional Tube
            \label{sec:bending_2D_tube}
        }
This section gives a detailed account of how an artificial  
two-dimensional tube can be unstable against bending,
{ i.e.}, bending may reduce the tube's free 
energy.  Although myelin coiling occurs in 
three dimensions, this two-dimensional 
case, nevertheless, captures all the essential features of 
the myelin coiling model  proposed in this work. 
Therefore, not only does this simple example offer a 
physical insight into why myelins coil, it also provides 
a framework for modeling myelin coiling later in this paper. 
In the following we will first define 
 the two-dimensional tube and specify how it deforms.
 Then  we will show that 
the tube may indeed curve to lower its energy.
The two-step deformation developed here makes it easy to 
determine changes in the energies.
%
%
\subsection{Geometry and Two-step Deformation
            \label{sec:2D_geometry}}
\begin{figure}
\begin{center}
\resizebox{0.75\columnwidth}{!}{
    \includegraphics{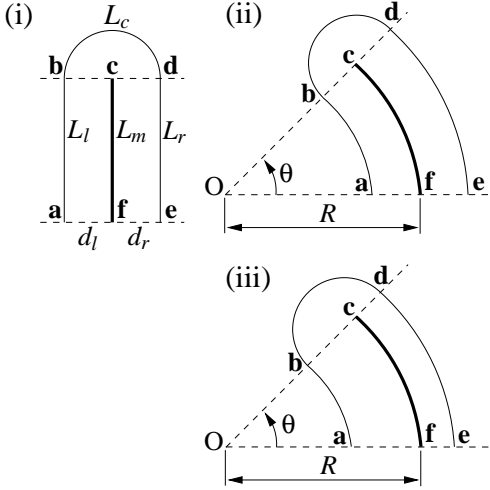}
    }
\end{center}
\caption{
    Two-step deformation of a two-dimensional tube. 
    The tube is initially in
    state (i) and transforms into the intermediate state
    (ii) and then into the final state (iii).
    During the deformation the tube area is conserved.
    The tube consists of a core, {\bf fc}, and 
    a perimeter, {\bf abde}. The core {\bf fc} is of zero 
    thickness and subject to bending only, { i.e.},
    its length is fixed. The perimeter {\bf abde}  is
    composed of two lines (arcs), {\bf ab} and {\bf de},
    and a half circle, {\bf bd}. As indicated in the figure, 
    the symbols $L_{\mbox{\scriptsize l}}$, $L_{\mbox{\scriptsize c}}$, $L_{\mbox{\scriptsize r}}$, $L_{\mbox{\scriptsize m}}$, $d_{\mbox{\scriptsize l}}$ and $d_{\mbox{\scriptsize r}}$ 
    denote the lengths of the lines (arcs) {\bf ab}, {\bf bd}, 
    {\bf de}, {\bf fc}, {\bf af} and {\bf fe}, respectively.
    Initial state (i): The tube is straight and its shape 
    is symmetric with respect to the line {\bf fc}, i.e.,
     $d_{\mbox{\scriptsize l}}=d_{\mbox{\scriptsize r}} \equiv d_0$, $L_{\mbox{\scriptsize l}}=L_{\mbox{\scriptsize r}}=L_{\mbox{\scriptsize m}} \equiv L_0$ and 
     thus $L_{\mbox{\scriptsize c}}=\pi d_0$. 
    Intermediate state (ii): The tube curves with
    $d_{\mbox{\scriptsize l}}=d_{\mbox{\scriptsize r}}=d_0$ and $L_{\mbox{\scriptsize m}} = R\theta = L_0$. 
    The point O is the center of bending 
    curvature, $R$ the radius of curvature of the 
    arc {\bf fc}, and $\theta$ the angle subtended 
    by {\bf fc}. The lengths $L_{\mbox{\scriptsize l}}=(R-d_0)\theta$, 
    $L_{\mbox{\scriptsize r}}=(R+d_0)\theta$, and $L_{\mbox{\scriptsize c}}=\pi d_0$.
    Final state (iii): The spacing $d_{\mbox{\scriptsize r}}$ decreases from 
    $d_0$ with fixed $R$ and $\theta$.
    In state (iii), $d_{\mbox{\scriptsize l}} > d_0 > d_{\mbox{\scriptsize r}}$, 
    $L_{\mbox{\scriptsize m}} = R\theta = L_0$, $L_{\mbox{\scriptsize l}}=(R-d_{\mbox{\scriptsize l}})\theta$, 
    $L_{\mbox{\scriptsize r}}=(R+d_{\mbox{\scriptsize r}})\theta$, and $L_{\mbox{\scriptsize c}}=\pi(d_{\mbox{\scriptsize l}}+d_{\mbox{\scriptsize r}})/2$. 
    Because of area conservation (\ref{eqn:A_2D}), 
    the width $(d_{\mbox{\scriptsize l}}+d_{\mbox{\scriptsize r}})$ of state (iii)
    is greater than the initial value $2d_0$.
\label{fig:2d_tube}
} 
\end{figure}
Figure \ref{fig:2d_tube} describes the geometry of
the two-dimensional tube and explains how it 
deforms. 
The tube comprises a core, {\bf fc}, and a perimeter, 
{\bf abde}, which are two-dimensional analogues of the 
inner and outer cylinders of the two-bilayer myelin studied
in Section \ref{sec:coiling_2_myelin}, respectively.
All the state parameters, such as $d_{\mbox{\scriptsize r}}$, $d_{\mbox{\scriptsize l}}$ and 
$\theta$, are defined in the figure caption.
The eccentricities of the bilayer cylinders in 
a myelin are simulated in this two-dimensional case 
by allowing the spacings $d_{\mbox{\scriptsize l}}$ and $d_{\mbox{\scriptsize r}}$  between the core 
and the perimeter to vary.  
The perimeter {\bf abde} of length
\begin{equation}
 L_{\mbox{\scriptsize 2D}} \equiv L_{\mbox{\scriptsize l}}+L_{\mbox{\scriptsize c}}+L_{\mbox{\scriptsize r}}
     \label{eqn:L_2D}
\end{equation}
is extensible and under uniform tension.
The core {\bf fc} exerts a force on the end cap {\bf bd} to 
counterbalance the tension of the perimeter and
prevent the tube from collapsing. 
For simplicity we assume the core  to be 
of zero thickness and subject to bending 
only, { i.e.}, its length $L_{\mbox{\scriptsize m}}$ is constant.
One may think of the core as a spring of infinite 
spring constant. Allowing $L_{\mbox{\scriptsize m}}$ to vary amounts to 
reducing the spring constant to a finite
value and thus adding one more degree of freedom 
to the system, which would not invalidate the conclusions 
of this section.

The tube area $A_{\mbox{\scriptsize 2D}}$ is postulated to be constant
(Fig. \ref{fig:2d_tube}):
\begin{eqnarray}
 A_{\mbox{\scriptsize 2D}} &=& 2 d_0 L_0 + \frac{\pi}{2} d_0^2 \cdot \hat{c}
              \nonumber \\
        &=& \frac{\theta}{2}\left[(R+d_{\mbox{\scriptsize r}})^2-(R-d_{\mbox{\scriptsize l}})^2\right]
         \nonumber \\
        & & \, +\; \frac{\pi}{2}\left( \frac{d_{\mbox{\scriptsize l}}+d_{\mbox{\scriptsize r}}}{2}\right)^2 
                    \hat{c},
            \label{eqn:A_2D}
\end{eqnarray}
where the angle $\theta$ can be larger than $2\pi$,
 { i.e.}, the tube may coil, and $\hat{c}=0 \mbox{ or } 1$. 
The  area conservation in two dimensions is analogous to 
the volume conservation in three dimensions 
(Sect. \ref{sec:modeling}).
In order to examine the effect of the end 
cap {\bf bd}, we deliberately multiply every end cap 
contribution by the parameter $\hat{c}$. 
We will demonstrate that when the tube's aspect ratio 
$L_0/2d_0$  is much greater than one, 
 all the end cap contributions are negligible, 
{ i.e.}, we can set $\hat{c}= 0$.

The deformation of this two-dimensional tube takes place
in two steps (Fig. \ref{fig:2d_tube}):
the tube is transformed from the initial state (i) into 
the intermediate state (ii) and then into the final state 
(iii).
The initial tube is straight and its shape is 
symmetric with respect to the line {\bf fc}. 
From state (i) to state (ii) the tube curves with fixed
$d_{\mbox{\scriptsize l}}$, $d_{\mbox{\scriptsize r}}$ and $L_{\mbox{\scriptsize m}}$. Consequently, the perimeter length
$L_{\mbox{\scriptsize 2D}}$ (\ref{eqn:L_2D}) is also unchanged.
In the final state (iii) the spacing $d_{\mbox{\scriptsize r}}$ decreases from 
its original value $d_0$ while $L_{\mbox{\scriptsize m}}$ is still unchanged.
Equation (\ref{eqn:A_2D}) gives
 the change in the perimeter length $L_{\mbox{\scriptsize 2D}}$ 
from state (i) or (ii) to state (iii) 
\begin{equation}
\delta L_{\mbox{\scriptsize 2D}} 
         =  \frac{L_0 \Delta_{\mbox{\scriptsize d}}}{R} (-1 + \frac{\pi d_0  \hat{c}}{2 L_0})
            + \cdots,
            \label{eqn:delta_L_2D}
\end{equation}
where $ \Delta_{\mbox{\scriptsize d}} \equiv d_{\mbox{\scriptsize l}}-d_{\mbox{\scriptsize r}}$. 
It is obvious that $\delta L_{\mbox{\scriptsize 2D}} < 0$ when $\Delta_{\mbox{\scriptsize d}} >0$ 
and $d_0 \ll L_0$. 
In state (iii) $\Delta_{\mbox{\scriptsize d}}$ has to be greater than zero 
because of  the area conservation (\ref{eqn:A_2D}).
We now investigate how
the tube's free energy changes with the parameters
$R$ and $\Delta_{\mbox{\scriptsize d}}$.
%
%
\subsection{Free Energy and Instability
            \label{sec:2D_energies}}
The total free energy  of the tube is given by
\begin{equation}
 F_{\mbox{\scriptsize 2D}} \equiv F_{\mbox{\scriptsize 2D}}^{\mbox{\scriptsize b}} + F_{\mbox{\scriptsize 2D}}^{\mbox{\scriptsize e}},
     \label{eqn:F_2D}
\end{equation}
where $F_{\mbox{\scriptsize 2D}}^{\mbox{\scriptsize b}}$ is the bending energy and $F_{\mbox{\scriptsize 2D}}^{\mbox{\scriptsize e}}$ 
the elastic energy associated with the tension 
in the perimeter.
The interaction between the core and the perimeter is 
neglected for the sake of simplicity as well as clarity. 
We assume that the bending energy $F_{\mbox{\scriptsize 2D}}^{\mbox{\scriptsize b}}$ takes
the  form of
\begin{equation}
    F_{\mbox{\scriptsize 2D}}^{\mbox{\scriptsize b}} = \frac{\kappa_{\mbox{\scriptsize 2D}}}{R^2}  L_0
                + a_0  \hat{c},
    \label{eqn:F_2D^b}
\end{equation}
where $\kappa_{\mbox{\scriptsize 2D}}$ is the bending stiffness of the tube,
$R$ the radius of bending curvature, and $L_0=R\theta$ the 
tube length (Fig. \ref{fig:2d_tube}). 
The last term $a_0\hat{c}$ represents the contribution
of the end cap {\bf bd}. 
The tension $\Sigma$ in the perimeter is defined as
\begin{equation}
  \Sigma \equiv \frac{\partial F_{\mbox{\scriptsize 2D}}^{\mbox{\scriptsize e}}}{\partial L_{\mbox{\scriptsize 2D}}},
      \label{eqn:Sigma}
\end{equation}
where $L_{\mbox{\scriptsize 2D}}$ is the perimeter length (\ref{eqn:L_2D}).
Now we can determine the difference in the total 
free energy $F_{\mbox{\scriptsize 2D}}$ between the initial state (i) and 
the final state (iii).

When the tube is deformed from state (i) to state (ii),
only the bending energy $F_{\mbox{\scriptsize 2D}}^{\mbox{\scriptsize b}}$ (\ref{eqn:F_2D^b})
changes because $d_{\mbox{\scriptsize l}}$, $d_{\mbox{\scriptsize r}}$, $L_{\mbox{\scriptsize m}}$ and thus $L_{\mbox{\scriptsize 2D}}$ are 
 unchanged (Fig. \ref{fig:2d_tube}). Therefore
\begin{equation}
  F^{\mbox{\scriptsize b}}_{\mbox{\scriptsize 2D}}[\mbox{(ii)}]- F^{\mbox{\scriptsize b}}_{\mbox{\scriptsize 2D}}[\mbox{(i)}] 
        =\frac{\kappa_{\mbox{\scriptsize 2D}}}{R^2}  L_0.
           \label{eqn:F_2D^b_i_ii}
\end{equation}
The change in $F_{\mbox{\scriptsize 2D}}^{\mbox{\scriptsize b}}$ from state (ii), 
where $\Delta_{\mbox{\scriptsize d}}=0$, to state (iii), where $\Delta_{\mbox{\scriptsize d}} >0$,
is evidently given by
\begin{equation}
    F_{\mbox{\scriptsize 2D}}^{\mbox{\scriptsize b}}[\mbox{(iii)}]- F_{\mbox{\scriptsize 2D}}^{\mbox{\scriptsize b}}[\mbox{(ii)}]
    = \frac{b_1(R)}{R} L_0 \Delta_{\mbox{\scriptsize d}} 
    + b_2  \hat{c} + {\cal O}(\Delta_{\mbox{\scriptsize d}}^2).
           \label{eqn:F_2D^b_ii_iii}
\end{equation}
The first two terms on the right-hand side of the 
above equation represent the lowest-order 
correction of the bending energy
due to a change in $\Delta_{\mbox{\scriptsize d}}$.
The term $b_2\hat{c}$, also a function of $\Delta_{\mbox{\scriptsize d}}$,
is the end cap contribution.
Combining (\ref{eqn:F_2D^b_i_ii}) and 
(\ref{eqn:F_2D^b_ii_iii}) leads to
\begin{eqnarray}
 \delta F_{\mbox{\scriptsize 2D}}^{\mbox{\scriptsize b}} & \equiv &  
         F_{\mbox{\scriptsize 2D}}^{\mbox{\scriptsize b}}[\mbox{(iii)}]- F_{\mbox{\scriptsize 2D}}^{\mbox{\scriptsize b}}[\mbox{(i)}]
           \nonumber \\
    &=& \frac{\kappa_{\mbox{\scriptsize 2D}}}{R^2}  L_0
        + \frac{b_1(R)}{R} L_0 \Delta_{\mbox{\scriptsize d}} 
        + b_2  \hat{c} + {\cal O}(\Delta_{\mbox{\scriptsize d}}^2)
         \label{eqn:delta_F_2D^b}\\
    & > &  0.
       \nonumber 
\end{eqnarray}
For cases with finite $R$, small $\Delta_{\mbox{\scriptsize d}}$, and 
large aspect ratio ($L_0\gg 2d_0$), equation
(\ref{eqn:F_2D^b_ii_iii}) is much less than equation
(\ref{eqn:F_2D^b_i_ii}) and thus can be neglected.
Equations (\ref{eqn:delta_L_2D}) and (\ref{eqn:Sigma}) give
 the change in the elastic energy $F_{\mbox{\scriptsize 2D}}^{\mbox{\scriptsize e}}$:
\begin{eqnarray}
  \delta F_{\mbox{\scriptsize 2D}}^{\mbox{\scriptsize e}} & \equiv &  
          F_{\mbox{\scriptsize 2D}}^{\mbox{\scriptsize e}}[\mbox{(iii)}]- F_{\mbox{\scriptsize 2D}}^{\mbox{\scriptsize e}}[\mbox{(i)}]
                         \nonumber \\
            & = & \Sigma \cdot \delta L_{\mbox{\scriptsize 2D}} + \frac{1}{2} 
            \frac{\partial \Sigma}{\partial L_{\mbox{\scriptsize 2D}}}
            (\delta L_{\mbox{\scriptsize 2D}})^2 + \cdots
                         \nonumber \\
            & = & \Sigma  \frac{L_0 \Delta_{\mbox{\scriptsize d}}}{R} 
                (-1 + \frac{\pi d_0 \hat{c}}{2 L_0} )
             + {\cal O}(\Delta_{\mbox{\scriptsize d}}^2).
                 \label{eqn:delta_F_2D^e}
\end{eqnarray}

Equations (\ref{eqn:delta_F_2D^b}) and (\ref{eqn:delta_F_2D^e})
lead to the change in $F_{\mbox{\scriptsize 2D}}$ (\ref{eqn:F_2D})
from state (i) to state (iii):
\begin{eqnarray}
    \delta F_{\mbox{\scriptsize 2D}}(R, \Delta_{\mbox{\scriptsize d}}) & = & \delta F_{\mbox{\scriptsize 2D}}^{\mbox{\scriptsize b}} + 
             \delta F_{\mbox{\scriptsize 2D}}^{\mbox{\scriptsize e}} 
             \nonumber \\
        & = & \frac{L_0}{R} \left[ \frac{\kappa_{\mbox{\scriptsize 2D}}}{R} 
            + \Delta_{\mbox{\scriptsize d}} \left( b_1 - \Sigma \right)
            \right]
        \nonumber \\
        & & +\; \hat{c} \left[ b_2 +
            \frac{\pi d_0 \Sigma \Delta_{\mbox{\scriptsize d}}}{2R}
        \right] + {\cal O}(\Delta_{\mbox{\scriptsize d}}^2)
       \label{eqn:delta_F_2D} \\
        & \rightarrow &  \frac{L_0}{R}\left[\frac{\kappa_{\mbox{\scriptsize 2D}}}{R} 
            - \Delta_{\mbox{\scriptsize d}} \,\Sigma \right],
        \nonumber
\end{eqnarray}
where $b_1 \Delta_{\mbox{\scriptsize d}}$ is much less than $\kappa_{\mbox{\scriptsize 2D}}/R$ 
and thus neglected
(see the text below (\ref{eqn:delta_F_2D^b})).
Although $\delta F_{\mbox{\scriptsize 2D}}^{\mbox{\scriptsize b}} \geq 0$, $\delta F_{\mbox{\scriptsize 2D}}^{\mbox{\scriptsize e}}$ 
 and hence $\delta F_{\mbox{\scriptsize 2D}}$ can be negative. Consequently, 
 if the perimeter tension $\Sigma$ is large enough,  
 a tube in state (i) would curve to lower its free energy.
 The equilibrium value of $(R, \Delta_{\mbox{\scriptsize d}})$ 
 can be determined if the ${\cal O}(\Delta_{\mbox{\scriptsize d}}^2)$ term in 
 (\ref{eqn:delta_F_2D}) is known.
In addition, all the above equations involving the end 
cap show that as long as the tube is sufficiently 
long, the end cap contributions are negligible, 
 { i.e.}, $\hat{c}$ can be set to zero.
%
%
\subsection{Summary}
We have shown that a two-dimensional 
tube of a given area is unstable against bending
when the perimeter tension  is sufficiently large. 
Owing to the area conservation, the perimeter length 
decreases  when the tube undergoes the two-step deformation 
(Fig. \ref{fig:2d_tube}). The decrease in the perimeter 
length reduces the elastic energy $F_{\mbox{\scriptsize 2D}}^{\mbox{\scriptsize e}}$ 
(\ref{eqn:Sigma}). 
Therefore, the tube can bend to lower its 
energy if the reduction in $F_{\mbox{\scriptsize 2D}}^{\mbox{\scriptsize e}}$ is 
larger than the increase in the bending energy $F_{\mbox{\scriptsize 2D}}^{\mbox{\scriptsize b}}$.
We also have shown that the end cap energy and its changes 
are negligible when the tube has a large aspect 
ratio, { i.e.}, $L_0/2d_0\gg 1$.

This two-dimensional case will 
serve as a paradigm for modeling myelin coiling 
in the following sections.
In addition to the bilayer bending rigidity and the 
bilayer lateral tension, the inter-bilayer interaction 
will also be taken into consideration in our myelin model.
We will employ  the two-step
deformation  (Fig. \ref{fig:2d_tube}) to determine
the energy changes associated with myelin coiling.
%
%
\section{Model Definition\label{sec:model}}
The myelin model proposed in this section takes the bilayer 
tension and cylinder eccentricity into account 
(Sect. \ref{sec:contact}).
We postulate that the growth of myelins is sufficiently slow
so that the myelins are in quasi-equilibrium and therefore
their energies are well defined 
\cite{Huang-2005,Santangelo-2002}. 
In the following we will first specify the geometry and 
energy of a myelin and
then  describe how myelin coiling is modeled in this 
work.
Like the two-dimensional case of the previous section,
the proposed coiling mechanism is analogous to the 
Euler buckling of an elastic rod subject to axial 
compression \cite{Love-1944}. 

%
%
\subsection{Myelin Geometry}
\begin{figure}
\begin{center}
\resizebox{0.6\columnwidth}{!}{
    \includegraphics{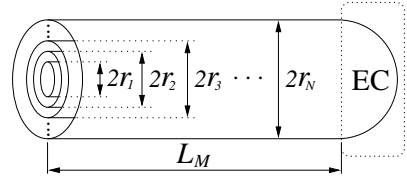}
    }
\end{center}
\caption{ 
     Model of an $N$-bilayer myelin. The myelin is 
     composed of $N$ nested bilayer cylinders. 
    These cylinders may not be coaxial and moreover,
    they can bend cooperatively to yield myelin
    coiling (Figs. \ref{fig:min_dist} and \ref{fig:coil}).
     The symbol $r_n$ denotes the radius of the $n$-th 
     cylinder, where $1\leq n \leq N$. 
    Since these bilayer cylinders need not 
     be concentric, we label their diameters $2 r_n$
     instead of their radii $r_n$.
     The length $L_n$ of the $n$-th bilayer 
     cylinder is defined as that of its axis
     (Fig. \ref{fig:bending_n_tube}),
     and the length $L_{\mbox{\scriptsize M}}$ of 
     a myelin as that of the innermost cylinder,
     { i.e.}, $L_{\mbox{\scriptsize M}} \equiv L_1$. 
     For a straight myelin, $L_n$ and $L_{\mbox{\scriptsize M}}$  are equal 
     for all $n$, whereas for a curved myelin, they may 
     not (Fig. \ref{fig:bending_myelin}). 
      In this work we only consider myelins of large 
      aspect ratios, { i.e.}, $L_{\mbox{\scriptsize M}}/2r_N\gg 1$. Therefore
     the end caps (EC) of myelins can be neglected.     
\label{fig:myelin}               
}
\end{figure}

\begin{figure}
\begin{center}
\resizebox{0.57\columnwidth}{!}{
    \includegraphics{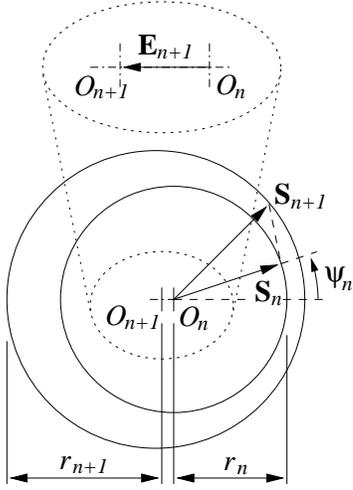}
    }
\end{center}
\caption{ 
 Cross-sectional view of two adjacent bilayer cylinders in 
 a myelin. 
We describe the $n$-th and $(n+1)$-th cylinders with the 
vectors $\mathbf{S}_n$ and $\mathbf{S}_{n+1}$, and 
their axes with $O_n$ and $O_{n+1}$, respectively.
When two nested cylinders are not concentric, 
the spacing between them is clearly not uniform
 (Fig. \ref{fig:bending_myelin}).
 In this work we define the spacing $d_n$ between the 
$n$-th and $(n+1)$-th cylinders as 
$d_n(\mathbf{S}_n)=\mbox{min}\{|\mathbf{S}_{n+1}-\mathbf{S}_n|\}$.
To calculate $d_n(\mathbf{S}_n)$, we select the coordinate
system with $O_n$ being the coordinate origin and  
$\mathbf{S}_n=(r_n \cos\psi_n, r_n \sin\psi_n)$.
The eccentricity vector $\mathbf{E}_{n+1}\equiv(\delta_{n+1}, 0)$ 
represents the positional difference between 
$O_{n+1}$ and $O_n$, as shown in the oval inset.
Given $\mathbf{\delta}_{n+1}$, 
the inter-bilayer spacing $d_n(\mathbf{S}_n)=
r_{n+1}-(r_n^2+\delta_{n+1}^2-2 r_n \delta_{n+1}\cos\psi_n)^{1/2}$.
For convenience we set $\delta_1\equiv 0$.
 \label{fig:min_dist}
}
\end{figure}

\begin{figure}
\begin{center}
\resizebox{0.47\columnwidth}{!}{
    \includegraphics{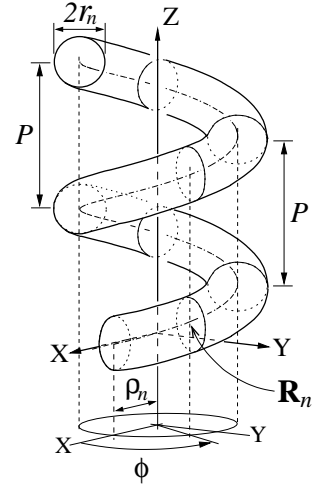}
    }
\end{center}
\caption{ 
     Coiling of a myelin. For clarity  
     we only show the $n$-th bilayer cylinder
     (Fig. \ref{fig:myelin}). 
    The axis of this coiled cylinder forms a 
    helix described by the parametric equation
    $\mathbf{R}_n(\phi) = (\rho_n \cos\phi, \rho_n \sin\phi,
    P\phi/2\pi)$, where $\rho_n$ is the radius of 
    the circular projection of the helix 
    on the X-Y plane and $P$ the pitch of the helix.
\label{fig:coil}
}
\end{figure}

\begin{figure}
\begin{center}
\resizebox{0.5\columnwidth}{!}{
    \includegraphics{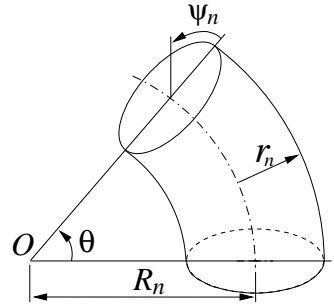}
    }
\end{center}
\caption{ 
 Bending of the $n$-th bilayer cylinder. 
 The point $O$ is the center of bending 
 curvature. The cylinder length $L_n \equiv R_n \theta$, 
 where $R_n$ is the radius of bending 
 curvature (\ref{eqn:R_n}) and 
 $\theta$ the angle subtended by the cylinder.
 This cylinder can be viewed as a section of a 
  torus of radii $r_n$ and $R_n$. 
 Straight cylinders correspond to cases with
 $R_n\rightarrow \infty$, $\theta\rightarrow 0$, and
 finite $R_n \theta$.
 The surface area element $dA_n=r_nd\psi_n(R_n+r_n\cos\psi_n)d\theta$.
 The two principal curvatures of the surface are $1/r_n$ and
 $\cos\psi_n/(R_n + r_n\cos\psi_n)$.
 \label{fig:bending_n_tube}
}
\end{figure}

An $N$-bilayer myelin is a 
multilamellar tube composed of a nested stack of 
 $N$ cylindrical fluid bilayers (Fig. \ref{fig:myelin}).
 These bilayer cylinders need not be coaxial 
 (Fig. \ref{fig:min_dist}) and
 may curve collectively to give rise to myelin 
 coiling (Fig. \ref{fig:coil}).
In addition, we only consider myelins of large aspect
ratios so that the end caps  are negligible. 
The symbol $r_n$ denotes the radius of the $n$-th
bilayer cylinder, with $n=1$ and $n=N$ corresponding to
the innermost and outermost cylinders, 
respectively (Fig. \ref{fig:myelin}).
We define the length $L_n$ of the $n$-th bilayer cylinder 
as that of its axis and the length $L_{\mbox{\scriptsize M}}$ of a myelin 
as that of the innermost cylinder, i.e.,  $L_{\mbox{\scriptsize M}}\equiv L_1$
(Figs. \ref{fig:myelin} and \ref{fig:bending_n_tube}).  
With this myelin model we can now 
describe the geometry of a coiled myelin.

A coiled myelin is composed of nested helical 
bilayer cylinders (Fig. \ref{fig:coil}).
The following parametric expression for a helix describes
the axis of the $n$-th  bilayer cylinder:
\begin{equation}
  \mathbf{R}_n (\phi)  = (\rho_n \cos\phi, \rho_n \sin\phi,
        P\phi/2\pi),
    \label{eqn:vec_R_n}
\end{equation}
where $\phi$, $\rho_n$ and $P$ are defined in
Figure \ref{fig:coil}. The radius of curvature of the helix
$\mathbf{R}_n$ is equal to
\begin{equation}
 R_n = \rho_n \left( 1+ \frac{P^2}{4\pi^2\rho_n^2} \right).
      \label{eqn:R_n}
\end{equation}
 The volume $V_n$, area $A_n $ and length $L_n$
 of the $n$-th bilayer cylinder are given by 
\begin{eqnarray}
    V_n  &=&  \pi r_n^2 L_n,
     \label{eqn:V_n} \\
    A_n  &=&  2 \pi r_n L_n,
     \label{eqn:A_n} \\
    L_n &=& 2\pi \rho_n \left[ 1+ \frac{P^2}{4\pi^2\rho_n^2} 
  \right]^{1/2} t_{\mbox{\scriptsize c}},
       \label{eqn:L_n}
\end{eqnarray}
where $r_n$ is the radius of the cylinder and
$t_{\mbox{\scriptsize c}}$ the number of turns of the coiling 
(Fig. \ref{fig:coil}).
Strictly speaking,  $A_n$ is the projected 
area of  the $n$-th bilayer \cite{Helfrich-1984}.
The true bilayer area is larger than $A_n$ because of 
 thermal undulations of the bilayer. 
%
%
\subsection{Free Energy}
The total energy $F$ of a myelin includes
the curvature energy $F^{\mbox{\scriptsize c}}$ of bilayers, the elastic 
energy $F^{\mbox{\scriptsize e}}$ associated with bilayer lateral tension,  
  and the inter-bilayer interaction energy $F^{\mbox{\scriptsize i}}$:
\begin{equation}
 F \equiv F^{\mbox{\scriptsize c}} + F^{\mbox{\scriptsize e}} + F^{\mbox{\scriptsize i}}.
     \label{eqn:F}
\end{equation}

The curvature energy $E^{\mbox{\scriptsize c}}$ of a membrane of 
area $A$ takes the form of 
\begin{equation}
E^{\mbox{\scriptsize c}} = \frac{\kappa}{2}\int_A\;dA (K_1 + K_2-K_0)^2,
    \label{eqn:E^c}
\end{equation}
where $\kappa$ is the bending stiffness, and $K_1$ and
$K_2$ are the principal curvatures \cite{TW-2004}. 
The spontaneous curvature $K_0$ is taken to be zero for
simplicity. (See the Discussion section.)
The Gaussian curvature term ($\sim \int dA\, K_1 K_2$) is 
neglected because the myelin coiling investigated here does 
not involve any topological changes. 
Thus, there can be no change in this energy according to 
the Gauss-Bonnet theorem \cite{Struik-1950}.
Because of the fluid nature of bilayers, a myelin can not 
sustain torsional stress in its axial direction.
Therefore the  curvature energy change 
of a myelin due to coiling  depends only on the bending 
of its constituent cylinders (Figs. \ref{fig:coil} and
\ref{fig:bending_n_tube}). 
Using (\ref{eqn:R_n}) and (\ref{eqn:E^c}), with
the area element $dA_n=r_nd\psi_n(R_n+r_n\cos\psi_n)d\theta$ 
and the two principal curvatures $1/r_n$ and 
$\cos\psi_n/(R_n + r_n\cos\psi_n)$ 
(Fig. \ref{fig:bending_n_tube}),
the curvature energy $E^{\mbox{\scriptsize c}}_n$ of the $n$-th bilayer 
cylinder in a coiled myelin is given by
\begin{eqnarray}
E^{\mbox{\scriptsize c}}_n & = & \frac{\pi\kappa L_n}{r_n\sqrt{1-(r_n/R_n)^2}}
        \nonumber \\
    & = & \frac{\pi\kappa L_n}{r_n} + \frac{\pi\kappa r_n L_n}{2 R_n^2}
        \left[1 + \frac{3}{4}\frac{r_n^2}{R_n^2} + \cdots
        \right],
    \label{eqn:E^c_n}
\end{eqnarray}
where  $L_n=R_n\theta$ is the cylinder length. 
The above equation suggests that the curvature energy of 
a curved bilayer cylinder can be written as a sum of 
the energy ($\sim 1/r_n$) associated with local 
bending of the bilayer 
and that ($\sim 1/R_n^2$) resulting from bending of 
the cylinder (\ref{eqn:R_n}).
Equation (\ref{eqn:E^c_n}) leads to the curvature energy
$F^{\mbox{\scriptsize c}}$ of a coiled $N$-bilayer myelin:
\begin{eqnarray}
    F^{\mbox{\scriptsize c}} &=& \sum_{n=1}^{N} E^{\mbox{\scriptsize c}}_n 
        \nonumber \\
    & = & \sum_{n=1}^{N} \frac{\pi\kappa L_n}{r_n}
         + \sum_{n=1}^{N} \frac{\tilde{\kappa}_n L_n}{2 R_n^2},
       \label{eqn:F^c_0}
\end{eqnarray}
where the bending modulus $\tilde{\kappa}_n$ of the 
$n$-th bilayer cylinder is defined as
\begin{equation}
\tilde{\kappa}_n \equiv \pi\kappa r_n
        \left[1 + \frac{3}{4}\frac{r_n^2}{R_n^2} + \cdots
        \right].
    \label{eqn:tilde_kappa_n}
\end{equation}

The lateral tension of a bilayer arises from
stretching the bilayer area from the relaxed 
state \cite{Helfrich-1984,Evans-1987,Evans-1990}.  
For a fluid bilayer, the tension is uniform and isotropic.  
Let $F_n^{\mbox{\scriptsize e}}$ denote the elastic
energy giving rise to the tension $\sigma_n$ in
the $n$-th bilayer cylinder, { i.e.}, 
\begin{equation}
 \sigma_n \equiv \frac{\partial F_n^{\mbox{\scriptsize e}}}{\partial A_n},
     \label{eqn:sigma_n}
\end{equation}
where $A_n$ is the area of the cylinder (\ref{eqn:A_n}).
The  elastic energy $F^{\mbox{\scriptsize e}}$ of an $N$-bilayer myelin
is then given by
\begin{equation}
  F^{\mbox{\scriptsize e}} = \sum_{n=1}^{N} F_n^{\mbox{\scriptsize e}}.
      \label{eqn:F^e}
\end{equation}

The energy $F^{\mbox{\scriptsize i}}$ describes the net result of all kinds of
 inter-bilayer interactions such as  van der Waals 
 attraction, electrostatic repulsion \cite{Israelachvili-1993}, 
 hydration pressure \cite{Rand-1989}, and the Helfrich 
repulsion \cite{Helfrich-1978}. 
For an $N$-bilayer myelin, $F^{\mbox{\scriptsize i}}$ can be 
conveniently expressed as 
\begin{equation}
 F^{\mbox{\scriptsize i}} = \sum_{n=1}^{N-1} F_n^{\mbox{\scriptsize i}}, 
 \label{eqn:F^i}
\end{equation}
where $F_n^{\mbox{\scriptsize i}}$ represents the energy of interaction 
between the $n$-th and $(n+1)$-th bilayers.
We expect that $F_n^{\mbox{\scriptsize i}}$ can be written as 
\begin{equation}
 F_n^{\mbox{\scriptsize i}} \equiv \int_{A_n} dA_n\, f_n^{\mbox{\scriptsize i}}\left( d_n(\mathbf{S}_n),
     \sigma_n, \sigma_{n+1} \right),
 \label{eqn:f_n^i}
\end{equation}
where  $f_n^{\mbox{\scriptsize i}}$ denotes {\em the interaction energy $F_n^{\mbox{\scriptsize \rm i}}$
 per unit area of the $n$-th bilayer} 
and $d_n(\mathbf{S}_n)$, 
defined in Figure \ref{fig:min_dist},
is the spacing between  the $n$-th and 
$(n+1)$-th bilayers. 
For simplicity we assume that $f_n^{\mbox{\scriptsize i}}$ depends only on 
the spacing $d_n$  and the tensions $\sigma_n$ and 
$\sigma_{n+1}$ (\ref{eqn:sigma_n}). 
This implies 
$f^{\mbox{\scriptsize i}}_n(d_n,\sigma_n,\sigma_{n+1})
= f^{\mbox{\scriptsize i}}_n(d_n,\sigma_{n+1},\sigma_{n})$
since the  effect of the bilayer 
curvatures is neglected \cite{Huang-2005}.
Because the bilayer cylinders may not be coaxial, 
we define the inter-bilayer 
spacing $d_n(\mathbf{S}_n)$ as the shortest distance 
from  the point $\mathbf{S}_n$ on the $n$-th cylinder to
the $(n+1)$-th cylinder:
With $\delta_{n+1}$ and $\psi_n$ defined in 
Figure \ref{fig:min_dist},
\begin{eqnarray}
d_n(\mathbf{S}_n) & \equiv & d_n(\psi_n,\delta_{n+1}) 
                 \nonumber \\
    &=& 
r_{n+1}-\left(r_n^2+\delta_{n+1}^2-2 r_n \delta_{n+1} \cos\psi_n \right)^{1/2}
    \nonumber \\
    &=& (r_{n+1}-r_n) + \delta_{n+1} \left[\cos\psi_n
        - \frac{\sin^2\psi_n}{2} \frac{\delta_{n+1}}{r_n}
        \right.
        \nonumber \\
    & &  \left. 
    - \frac{\cos\psi_n \sin^2\psi_n}{2} \frac{\delta_{n+1}^2}{r_n^2}
        +\; {\cal O}\left(\frac{\delta_{n+1}^3}{r_n^3}\right) \right].
    \label{eqn:d_n}
\end{eqnarray}

The potential $f^{\mbox{\scriptsize i}}_n$ gives rise to an inter-bilayer pressure
\begin{equation}
    p_n \equiv - \frac{\partial f^{\mbox{\scriptsize i}}_n}{\partial d_n}.
    \label{eqn:p_n}
\end{equation}
This pressure may not be equal to the real inter-bilayer 
pressure ${\cal P}_n$ owing to the fact that
the definitions of  $f^{\mbox{\scriptsize i}}_n$ and $d_n$ are artificial. 
Nonetheless,  $p_n$ should be a good 
approximation to ${\cal P}_n$,
especially when $|\delta_{n+1}| \ll (r_{n+1}-r_n)$.
For typical myelins, we expect $p_n > 0$ \cite{Huang-2005}
and moreover, $p_n$ should satisfy
the force balance equation for 
the $n$-th bilayer (Laplace's formula) \cite{Diamant-2001}:
\begin{equation}
 \sigma_n \approx
 \frac{dp^{\mbox{\scriptsize w}}_n + p_{n-1}-p_{n}}{K_{n1}+K_{n2}},
    \label{eqn:force_n}
\end{equation}
where $dp^{\mbox{\scriptsize w}}_n$ is the water pressure 
difference across  the  bilayer, and
$K_{n1}$ and $K_{n2}$ are the two principal 
curvatures of the bilayer 
(see (\ref{eqn:E^c}) and Fig. \ref{fig:bending_n_tube}).

%
%
\subsection{Modeling Myelin Coiling
                \label{sec:modeling}
            }
The myelin structure defined in this work has three
important properties: 
(a) it is postulated to be in quasi-equilibrium owing to 
 its slow growth rate (Sect. \ref{sec:contact})
 \cite{Huang-2005,Santangelo-2002},
(b) the bilayer tension may vary from bilayer to bilayer,
and
(c) the constituent bilayer cylinders can be non-coaxial. 
Given these properties, we will demonstrate 
 that coiling may lead to a decrease in 
the elastic energy $F^{\mbox{\scriptsize e}}$ (\ref{eqn:F^e}).
Therefore, a straight myelin is unstable to coiling
if the decrease in $F^{\mbox{\scriptsize e}}$ is large
 enough to compensate for the net increase in other energies
 (see (\ref{eqn:F}), (\ref{eqn:F^c_0}), and (\ref{eqn:F^i})). 
In Section \ref{sec:coiling_N_myelin} we will derive
the energy changes associated with myelin coiling.
By analyzing a two-bilayer case in detail 
in Section \ref{sec:coiling_2_myelin},
we will show that coiling may indeed lower the myelin energy.

\begin{figure}
\begin{center}
\resizebox{0.75\columnwidth}{!}{
    \includegraphics{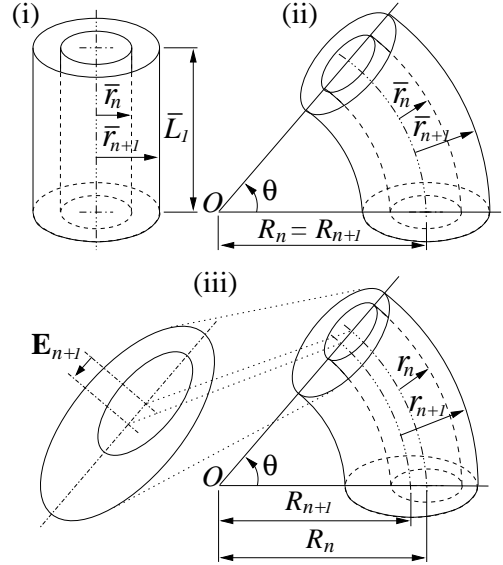}
    }
\end{center}
\caption{ 
    Two-step deformation of a myelin. 
    As in Figure \ref{fig:2d_tube}, this procedure
    transforms a myelin from state (i) into state (ii) 
    and then into state (iii), with the constraints of 
    fixed myelin length and fixed volume for each 
    bilayer cylinder. 
    For clarity we only show two adjacent bilayer cylinders
    (Fig. \ref{fig:min_dist}). 
    Initial state (i): The myelin consists of straight,
    coaxial bilayer cylinders with radii
    $r_n=\bar{r}_n$ and myelin
    length $L_{\mbox{\scriptsize M}}\equiv L_1 = \bar{L}_1=\bar{L}_n$
    (Fig. \ref{fig:myelin}).
    Intermediate state (ii): The myelin curves
    while its constituent cylinders are still coaxial. 
    In this state $R_n=R_1$.
    The constraints yield that  $R_n\theta = \bar{L}_1$,
    $r_n=\bar{r}_n$, and therefore 
    $A_n=\bar{A}_n$ (see (\ref{eqn:V_n}) and (\ref{eqn:A_n})).
    Final state (iii): 
    The cylinders are no longer coaxial.
    The eccentricity vector $\mathbf{E}_{n+1}$ is defined in 
    Figure \ref{fig:min_dist}.
    All the cylinders except for the innermost one 
    ($n=1$) may move toward the center $O$ of bending 
    curvature, { i.e.}, $\delta_{n+1}\leq 0$.
    In state (iii), $r_1=\bar{r}_1$, $L_1=\bar{L}_1$,
    $R_{n+1}=R_n+\delta_{n+1} \leq R_n$, and 
    therefore $L_{n+1}=R_{n+1}\theta \leq L_n=R_n\theta$.
    The constraints lead to $r_n\geq \bar{r}_n$ and 
    $A_n\leq \bar{A}_n$ for $n>1$ 
    (see (\ref{eqn:r_n}) and (\ref{eqn:A_n_iii})).
\label{fig:bending_myelin}
}
\end{figure}

We will employ  the two-step deformation 
developed in Section \ref{sec:bending_2D_tube}  to 
investigate the energy changes due to myelin coiling
(Figs. \ref{fig:2d_tube}  and \ref{fig:bending_myelin}).
For the sake of simplicity, two constraints are imposed.
First, the myelin length $L_{\mbox{\scriptsize M}}$ is fixed during the 
deformation (Fig. \ref{fig:myelin}). 
As in the two-dimensional case 
(cf. the text below (\ref{eqn:L_2D})), 
this artificial constraint would not invalidate our model.
Secondly, we do not allow  exchange of material, 
 surfactant or water, between bilayers. 
 Because water and bilayers are virtually incompressible 
under moderate pressure \cite{Rand-1989}, this constraint 
implies that the volume of each bilayer cylinder is 
conserved during the deformation. 
(See the Discussion section.)

Figure \ref{fig:bending_myelin} gives a detailed description 
of the two-step deformation. The myelin is deformed 
from state (i) into state (ii) and then into state (iii).
In state (i) the myelin is straight and its constituent 
bilayer cylinders are concentric, { i.e.}, 
$\mathbf{E}_n=\mathbf{0}$ for all $n$. 
From (i) to (ii), the myelin curves with the 
condition that the bilayer cylinders are still concentric. 
As a result, the curvature energy $F^{\mbox{\scriptsize c}}$ increases while
the elastic energy $F^{\mbox{\scriptsize e}}$ and the inter-bilayer 
interaction $F^{\mbox{\scriptsize i}}$ remain unchanged.
From (ii) to (iii), the cylinders are displaced so that 
 they are no longer coaxial (Fig. \ref{fig:min_dist}). 
In this step the displacement $\delta_n$ may become negative
 for $n>1$, reducing the bilayer area $A_n$ and thus the 
 elastic energy $F^{\mbox{\scriptsize e}}$.

Hereinafter, symbols like $\bar{r}_n$, $\bar{L}_n$, and
$\bar{p}_n$ with an
overhead bar refer to the initial state (i),
where the myelin is straight, and those without refer
to the final state (iii),  where the myelin is curved 
(Fig. \ref{fig:bending_myelin}).
The terms ``coiling'' and ``bending'' are used
interchangeably because they are energetically equivalent
for myelins composed of fluid bilayers.
Since the myelins considered here have large aspect ratios,
 their end caps are neglected. 
However, the end caps can be neglected only in terms of 
energy calculation. Their presence is necessary for 
volume conservation and force cancellation
(cf. text below (\ref{eqn:L_2D})).

%
%
\section{Bending of a Myelin\label{sec:coiling_N_myelin}
        }
Now we will use the  model defined in the previous 
section to investigate myelin coiling.
As in Section \ref{sec:bending_2D_tube}, 
we will first derive the geometrical changes of a myelin
undergoing the two-step deformation 
(Fig. \ref{fig:bending_myelin}).
Then we will determine the respective energy changes.
 For a myelin composed of $N$ bilayers, state (iii) is
 described by $N$ parameters,
 $R_1$, $\delta_2$, \dots, $\delta_{N-1}$, and $\delta_N$
($\delta_1\equiv 0$; see Fig. \ref{fig:min_dist}).

The two-step deformation is constrained by
the conditions of fixed myelin length 
$L_{\mbox{\scriptsize M}}\equiv L_1=\bar{L}_1=\bar{L}_n$ 
and volume conservation $V_n=\bar{V}_n$
for each bilayer cylinder (Sect. \ref{sec:modeling}).
With these two constraints and (Figs. \ref{fig:myelin},
\ref{fig:min_dist}, and \ref{fig:bending_myelin})
\[
R_n = R_1 + \sum_{j=1}^n \delta_j,
\]
the length $L_n$, radius $r_n$, and area $A_n$ of 
the $n$-th bilayer cylinder in state (iii) are given by  
((\ref{eqn:V_n}) and (\ref{eqn:A_n}))
\begin{eqnarray}
L_n &\equiv& R_n\theta =\bar{L}_n\left(1 + \sum_{j=1}^n \frac{\delta_j}{R_1}\right)
    \leq \bar{L}_n,
    \label{eqn:L_n_1} \\
r_n &=& \bar{r}_n \left(1 + \sum_{j=1}^n \frac{\delta_j}{R_1}\right)^{-1/2}
    \geq \; \bar{r}_n, 
    \label{eqn:r_n} \\
A_n &=& \bar{A}_n \left(1 + \sum_{j=1}^n \frac{\delta_j}{R_1} \right)^{1/2}
    \leq \; \bar{A}_n,
    \label{eqn:A_n_iii}
\end{eqnarray}
where $\delta_1 = 0$ by construction 
and $\delta_n \leq 0$ for $n>1$. 
Equation (\ref{eqn:r_n}) implies that not only $r_{n+1} \geq \bar{r}_{n+1}$
but also $(r_{n+1}-r_n) \geq (\bar{r}_{n+1}-\bar{r}_n)$, 
{ i.e.}, 
both $r_{n+1}$ and $(r_{n+1}-r_n)$  increase when the
myelin is transformed from state (i) to state (iii).
The increase in $(r_{n+1}-r_n)$ hints the possibility of
a decrease in the inter-bilayer interaction $F^{\mbox{\scriptsize i}}_n$ 
(\ref{eqn:f_n^i}). 
Merging (\ref{eqn:d_n}) and (\ref{eqn:r_n}) leads to the 
change in the inter-bilayer spacing 
\begin{eqnarray}
    \delta d_n &\equiv & d_n[\mbox{(iii)}]-d_n[\mbox{(i)}]
       \nonumber \\
    &=& d_n(\mathbf{S}_n) - (\bar{r}_{n+1}-\bar{r}_n)
        \nonumber \\
    &=& \delta d_{n1} + \delta d_{n2},
       \label{eqn:delta_d_n}
\end{eqnarray}
where
\begin{eqnarray}
    \delta d_{n1} & = &  -  \frac{\bar{r}_{n+1}}{2} \frac{\delta_{n+1}}{R_1}
        -\frac{(\bar{r}_{n+1}-\bar{r}_n)}{2} \sum_{j=1}^n \frac{\delta_j}{R_1}
        \nonumber \\
    & &  +\; \frac{3 \bar{r}_{n+1}}{8} \left(\sum_{j=1}^{n+1}\frac{\delta_j}{R_1}\right)^2 
        - \frac{3 \bar{r}_{n}}{8} \left(\sum_{j=1}^{n}\frac{\delta_j}{R_1}\right)^2 
        \nonumber \\
    & & + \;\; {\cal O}\left(\frac{\delta_i \delta_j\delta_k}{R_1^3}\right)
        \nonumber
\end{eqnarray}
and 
\begin{eqnarray}
 \delta d_{n2} & = & \; \delta_{n+1} \left[\cos\psi_n
        - \frac{\sin^2\psi_n}{2} \frac{\delta_{n+1}}{\bar{r}_n}
        \right.
        \nonumber \\
    & &  \left. \;\;\;\;\;\;\;\;\;\;\;\;
    - \frac{\cos\psi_n \sin^2\psi_n}{2} \frac{\delta_{n+1}^2}{\bar{r}_n^2}
    \;+ \;\cdots \; \right].
        \nonumber
\end{eqnarray}
Equation (\ref{eqn:A_n_iii}) gives rise to the area change 
\begin{eqnarray}
\delta A_n &\equiv & A_n[\mbox{(iii)}]-A_n[\mbox{(i)}]
        \nonumber \\
    &=&  \frac{\bar{A}_n}{2} \sum_{j=1}^n \frac{\delta_j}{R_1}
        - \frac{\bar{A}_n}{8} \left(\sum_{j=1}^n \frac{\delta_j}{R_1}\right)^2
     \nonumber \\
    & & + \; {\cal O}\left(\frac{\delta_i \delta_j\delta_k}{R_1^3}\right)
         \; \leq \; 0,
        \label{eqn:delta_A_n}
\end{eqnarray}
where $\delta_1=0$ and thus $\delta A_1=0$.
Equations (\ref{eqn:L_n_1})--(\ref{eqn:delta_A_n}) 
allow us to calculate the energy changes due to 
the two-step deformation, as shown below.

The curvature energy change (cf. (\ref{eqn:F^c_0}))
\begin{eqnarray}
\delta F^{\mbox{\scriptsize c}} &\equiv & F^{\mbox{\scriptsize c}}[\mbox{(iii)}] -F^{\mbox{\scriptsize c}}[\mbox{(i)}]
    \nonumber \\
    &=& \sum_{n=1}^N \frac{\bar{A}_n}{2} \left[
    \frac{\bar{\tilde{\kappa}}_n }{2\pi\bar{r}_n R_1^2}
        + g_n \sum_{j=1}^{n}\frac{\delta_j}{R_1}\right]
        \nonumber   \\
    &  & \;+\; {\cal O}\left(\frac{\delta_i \delta_j}{R_1^2}\right) > 0, 
       \label{eqn:delta_F^c_N}
\end{eqnarray}
where $\bar{L}_n=\bar{A}_n/2\pi\bar{r}_n=\bar{L}_1$ 
and (cf. (\ref{eqn:tilde_kappa_n}))
\begin{equation}
\bar{\tilde{\kappa}}_n \equiv \left(\tilde{\kappa}_n\right)_{\,\mbox{i}}
    = \pi\kappa \bar{r}_n
    \left[1 + \frac{3}{4}\frac{\bar{r}_n^2}{R_1^2} + \cdots
        \right].
    \label{eqn:bar_tilde_kappa_n}
\end{equation}
The first-order terms ($\sim g_n\sum \delta_j$) 
in (\ref{eqn:delta_F^c_N})
result from the second step of the deformation. 
The exact functional form of $g_n$ is  
unimportant because the lowest-order terms 
($\sim \bar{\tilde{\kappa}}_n$) dominate. 
This is also intuitively reasonable in that 
when $\sum\delta_n \ll R_1 < \infty$, 
the major contribution to $\delta F^{\mbox{\scriptsize c}}$ should come
from the deformation from (i) to  (ii) 
(Fig. \ref{fig:bending_myelin}). 
Therefore we will not derive $g_n$ here.

Equation \ref{eqn:delta_A_n} shows that for $n>1$,
 the area $A_n$ decreases in the two-step deformation. 
This in turn reduces the elastic energy $F^{\mbox{\scriptsize e}}$ 
(see (\ref{eqn:sigma_n}) and (\ref{eqn:F^e})).
With (\ref{eqn:delta_A_n}) the change in $F^{\mbox{\scriptsize e}}$ is given by 
\begin{eqnarray}
\delta F^{\mbox{\scriptsize e}} &\equiv & F^{\mbox{\scriptsize e}}[\mbox{(iii)}] -F^{\mbox{\scriptsize e}}[\mbox{(i)}] 
                = \sum_{n=1}^N \delta F_n^{\mbox{\scriptsize e}}
            \nonumber \\
    &=& \sum_{n=1}^N \left[\left(\frac{\partial F_n^{\mbox{\scriptsize e}}}{\partial A_n}\right)_{\mbox{i}} \delta A_n
    +\frac{1}{2} \left(\frac{\partial^2 F_n^{\mbox{\scriptsize e}}}{\partial A_n^2}\right)_{\mbox{i}} \delta A_n^2
 + \cdots \right]
    \nonumber \\
    &=& \sum_{n=1}^N \frac{\bar{A}_n \bar{\sigma}_n}{2} \sum_{j=1}^n \frac{\delta_j}{R_1}
    \nonumber \\
    & & + \sum_{n=1}^N \frac{\bar{A}_n}{8}
        \left[ \left(\frac{\partial \sigma_n}{\partial A_n}\right)_{\mbox{i}}\bar{A}_n
                -\bar{\sigma}_n \right]
        \left(\sum_{j=1}^{n}\frac{\delta_j}{R_1}\right)^2
    \nonumber \\
    & & + \;{\cal O}\left(\frac{\delta_i \delta_j\delta_k}{R_1^3}\right),
    \label{eqn:delta_F^e_N}
\end{eqnarray}
where $\delta F_1^{\mbox{\scriptsize e}} =0$ since $\delta A_1=0$.

The change in the inter-bilayer interaction $F^{\mbox{\scriptsize i}}$ 
(\ref{eqn:F^i})
takes a complex form because $\delta A_n \neq 0$ and
$\delta \sigma_n\neq 0$ for $n>1$:
\begin{eqnarray}
 \delta F^{\mbox{\scriptsize i}} &\equiv & F^{\mbox{\scriptsize i}}[\mbox{(iii)}] -F^{\mbox{\scriptsize i}}[\mbox{(i)}]
               = \sum_{n=1}^{N-1} \delta F_n^{\mbox{\scriptsize i}}
    \nonumber \\
 &=& \sum_{n=1}^{N-1} \delta \int_{A_n}dA_n \; f_n^{\mbox{\scriptsize i}}(d_n(\mathbf{S}_n), \sigma_n, \sigma_{n+1}),
         \nonumber
\end{eqnarray}
where $f^{\mbox{\scriptsize i}}_n$ is defined in (\ref{eqn:f_n^i}).
Expanding the above equation around state (i) results in 
\begin{eqnarray}
 \delta F^{\mbox{\scriptsize i}} &=&  \sum_{n=1}^{N-1} \int_{\bar{A}_n}d\bar{A}_n \left[-\bar{p}_n \delta d_n
       - \left(\frac{\partial p_n}{\partial d_n} \right)_{\mbox{i}} \frac{\delta d_n^2}{2} \right.
         \nonumber \\
 & & \;\;\;\;\;  \left. + \left(\frac{\partial f_n^{\mbox{\scriptsize i}}}{\partial \sigma_n} \right)_{\mbox{i}} \delta \sigma_n 
      + \left(\frac{\partial f_n^{\mbox{\scriptsize i}}}{\partial \sigma_{n+1}} \right)_{\mbox{i}} \delta \sigma_{n+1}
      \right] 
          \nonumber \\
 & &  + \; \sum_{n=1}^{N-1} \delta A_n \; \bar{f}_n^{\mbox{\scriptsize i}}
         + \cdots,
    \label{eqn:delta_F^i_N0}
\end{eqnarray}
where the area element (Fig. \ref{fig:bending_n_tube}) 
\[
  d\bar{A}_n = R_1 d\theta \cdot \bar{r}_n  d\psi_n 
         \left(1+\frac{\bar{r}_n}{R_1}\cos\psi_n\right),
\]
the inter-bilayer pressure of state (i) 
(cf. (\ref{eqn:p_n}))
\begin{equation}
 \bar{p}_n = - \left(\frac{\partial f_n^{\mbox{\scriptsize i}}}{\partial d_n} \right)_{\mbox{i}},
    \label{eqn:bar_p_n}
\end{equation}    
and the spacing change $\delta d_n$ is given by
 (\ref{eqn:delta_d_n}).
 Using (\ref{eqn:delta_A_n}) and
\begin{equation}
\delta \sigma_n = \left(\frac{\partial \sigma_n}{\partial A_n}\right)_{\mbox{i}} \delta A_n
    + \left(\frac{\partial^2 \sigma_n}{\partial A_n^2}\right)_{\mbox{i}} \frac{\delta A_n^2}{2}
    +\cdots,
    \label{eqn:delta_sigma_n}
\end{equation}
 we determine the change in the inter-bilayer interaction 
 energy as well as the net change in the myelin energy 
((\ref{eqn:F}), (\ref{eqn:delta_F^c_N}), and 
(\ref{eqn:delta_F^e_N})):
\begin{eqnarray}
\delta F^{\mbox{\scriptsize i}} &=& \sum_{n=1}^{N-1}\frac{\bar{A}_n}{2} 
        \Biggl[\;
    \sum_{k=2}^3{\cal C}_{nk} \sum_{j=1}^{n+1}\frac{\delta_j}{R_1} 
           + {\cal C}_{n4} \sum_{j=1}^n \frac{\delta_j}{R_1}
        \nonumber \\
        & & \;\;\;\;\;\;\;\;\;\;\;\;\;\;\;\;
        + \,\;{\cal C}_{n5}\, \delta_{n+1}^2 \Biggr]
       + {\cal O}\left(\frac{\delta_i \delta_j}{R_1^2}, \delta_{n+1}^4\right)
       \label{eqn:delta_F^i_N}
\end{eqnarray}
and
\begin{eqnarray}
 \delta F  &\equiv & F[\mbox{(iii)}]- F[\mbox{(i)}]
     \,=\, \delta F^{\mbox{\scriptsize c}} + \delta F^{\mbox{\scriptsize e}} + \delta F^{\mbox{\scriptsize i}}
             \nonumber \\
    &=& \sum_{n=1}^{N}\frac{\bar{A}_n}{2}\left[{\cal B}_n +
    \sum_{k=0}^1{\cal C}_{nk} \sum_{j=1}^n \frac{\delta_j}{R_1} \right] 
        \nonumber \\
    & &  + \sum_{n=1}^{N-1}\frac{\bar{A}_n}{2}
        \Biggl[\;
    \sum_{k=2}^3{\cal C}_{nk}\sum_{j=1}^{n+1} \frac{\delta_j}{R_1}        
    +  {\cal C}_{n4} \sum_{j=1}^n \frac{\delta_j}{R_1}
        \nonumber \\
     & &  \;\;\;\;\;\;\;\;\;\;\;\;\;\;\;\;
         +\,\; {\cal C}_{n5} \, \delta_{n+1}^2  
        \Biggr]
     + {\cal O}\left(\frac{\delta_i \delta_j}{R_1^2}, \delta_{n+1}^4\right),
    \label{eqn:delta_F_N}
\end{eqnarray}
where 
\begin{eqnarray}
{\cal B}_n &=&\frac{\bar{\tilde{\kappa}}_n }{2\pi\bar{r}_n R_1^2}
        \approx \frac{\kappa}{2 R_1^2}\;>\; 0,
     \label{eqn:B_n} \\
 {\cal C}_{n0}&=& g_n,
     \nonumber \\
    {\cal C}_{n1} &=&  \bar{\sigma}_n,
    \nonumber \\
{\cal C}_{n2} &=& (\bar{r}_{n+1}-\bar{r}_n) \bar{p}_n,
        \nonumber \\
 {\cal C}_{n3} &=&  \left(\frac{\partial f_n^{\mbox{\scriptsize i}}}{\partial \sigma_{n+1}} \right)_{\mbox{i}}
       \left(\frac{\partial \sigma_{n+1}}{\partial A_{n+1}} \right)_{\mbox{i}}
        \bar{A}_{n+1},
      \nonumber \\
{\cal C}_{n4} &=& \left(\frac{\partial f_n^{\mbox{\scriptsize i}}}{\partial \sigma_n} \right)_{\mbox{i}}
       \left(\frac{\partial \sigma_n}{\partial A_n} \right)_{\mbox{i}}
        \bar{A}_n + \bar{f}_n^{\mbox{\scriptsize i}}
        \nonumber \\
    &=& \left[\frac{\partial \left(f_n^{\mbox{\scriptsize i}} A_n\right)}{\partial A_n} \right]_{\mbox{i}}
        = \left(\frac{\partial F_n^{\mbox{\scriptsize i}}}{\partial A_n} \right)_{\mbox{i}},
    \label{eqn:C_n4} \\
 {\cal C}_{n5} &=& - \frac{1}{2} \left(\frac{\partial p_n}{\partial d_n} \right)_{\mbox{i}},
\end{eqnarray}
$\delta_1=0$ by definition, and $\delta_n\leq 0$ for $n>1$.
In spite of the fact that equation (\ref{eqn:delta_F_N})
allows $\delta_n$ to be positive or negative, we 
are only interested in  the cases where $\delta_n \leq 0$
(Figs. \ref{fig:min_dist} and \ref{fig:bending_myelin}).
The terms ${\cal B}_n$ and ${\cal C}_{n0}$ 
 come from the curvature energy $F^{\mbox{\scriptsize c}}$ (\ref{eqn:F^c_0}),
${\cal C}_{n1}$ from the elastic energy $F^{\mbox{\scriptsize e}}$ (\ref{eqn:F^e}), 
and ${\cal C}_{n2} - {\cal C}_{n5}$ from the inter-bilayer 
interaction energy $F^{\mbox{\scriptsize i}}$ (\ref{eqn:F^i}).
As mentioned in the text below (\ref{eqn:bar_tilde_kappa_n}),
${\cal C}_{n0}$ is negligible.
In addition, we expect $p_n > 0$ and 
$(\partial p_n/\partial d_n) <  0$
for  myelins \cite{Huang-2005}.
Equation (\ref{eqn:delta_F_N}) seems too complicated to 
be useful for investigating myelin coiling: 
there are $2N$ initial conditions, 
$\bar{\sigma}_n$ and $\bar{r}_n$, and
$N$ independent parameters, $R_1$ and $\delta_n$ 
($\delta_1\equiv 0$; see Fig. \ref{fig:min_dist}).
However, this equation is not as worthless as it seems. 
In order for a myelin to coil, $\delta F$ 
(\ref{eqn:delta_F_N}) has to be less than zero.
The two-dimensional case of Section \ref{sec:bending_2D_tube}
suggests that if some of the bilayer tensions 
$\bar{\sigma}_n$ are sufficiently large,
the decrease in $F^{\mbox{\scriptsize e}}$ (\ref{eqn:delta_F^e_N}) 
might be large enough to yield a negative $\delta F$.
In the next section we will show that the proposed 
mechanism of myelin coiling can be easily understood
with the special case of $N=2$. 
The enormous number of parameters in equation 
(\ref{eqn:delta_F_N}) merely reflects  
the complexity of myelinic structures: for instance,
 two apparently similar myelins may have very different 
 bilayer tension profiles and therefore one may coil and 
 the other may not.
%
%
\section{Bending of a Two-bilayer Myelin
            \label{sec:coiling_2_myelin}
        }
In this section we inspect closely a special case of myelin
coiling, where the myelin is composed of only two 
bilayers, { i.e.}, $N=2$. 
One may think of this two-bilayer myelin as
a generalization of the two-dimensional tube of
Section \ref{sec:bending_2D_tube}.
Besides,  equation (\ref{eqn:delta_F_N}) suggests that 
in terms of energy calculation, an $N$-bilayer 
myelin can be viewed as a multilamellar tube composed of 
$(N-1)$ nested two-bilayer myelins.
Therefore  the results of this two-bilayer
case should be applicable to multiple-bilayer myelins.
In the following
we will first derive a simple criterion of coiling 
instability for a two-bilayer myelin 
and study a numerical example in detail 
in Sections \ref{sec:criterion} and \ref{sec:validity}.
Then in Section \ref{sec:generalization} 
 we will show how this two-bilayer example can be 
 generalized for myelins of many bilayers

For a two-bilayer myelin, two independent parameters
$R_1$ and $\delta_2$
determine state (iii) (Figs. \ref{fig:min_dist} and 
\ref{fig:bending_myelin}).
Equation (\ref{eqn:delta_F_N}) can be rewritten as
\begin{eqnarray}
 \delta F &\equiv& F[\mbox{(iii)}]- F[\mbox{(i)}]
         \nonumber \\
     &=&   \frac{\bar{A}_1 }{2}\left[ 
    {\cal B} + \frac{\delta_2}{R_1} \sum_{k=0}^3{\cal C}_k  
            +  {\cal C}_5\, \delta_2^2\right]
        \nonumber \\
    & &\, + \;{\cal O}\left(\frac{\delta_2^2}{R_1^2}, \delta_2^4\right),
    \label{eqn:delta_F_2}
\end{eqnarray}
where $\delta_2 \leq 0$ (Figs. \ref{fig:min_dist} and \ref{fig:bending_myelin}),
\begin{eqnarray}
 {\cal B} & = & 
         {\cal B}_1 + {\cal B}_2 \frac{\bar{A}_2}{\bar{A}_1}
         = \frac{\bar{\tilde{\kappa}}_1+\bar{\tilde{\kappa}}_2}{2\pi \bar{r}_1 R_1^2}> 0,
             \nonumber \\
  {\cal C}_0 & = & 
              {\cal C}_{20} \frac{\bar{A}_2}{\bar{A}_1}
            \,=\, g_2 \frac{\bar{r}_2}{\bar{r_1}}, 
            \nonumber \\
   {\cal C}_1 &=& 
             {\cal C}_{21} \frac{\bar{A}_2}{\bar{A}_1}
               \,=\, \bar{\sigma}_2 \frac{\bar{r}_2}{\bar{r_1}} > 0,
           \label{eqn:C_1} \\
   {\cal C}_2 &=& {\cal C}_{12} \,=\, (\bar{r}_2-\bar{r}_1) \bar{p}_1 > 0,
            \label{eqn:C_2} \\
    {\cal C}_3 &=& {\cal C}_{13} = 
        \left(\frac{\partial f_1^{\mbox{\scriptsize i}}}{\partial \sigma_2} \right)_{\mbox{i}}
       \left(\frac{\partial \sigma_2}{\partial A_2} \right)_{\mbox{i}}
      \bar{A}_2 < 0,
            \label{eqn:C_3} \\
    {\cal C}_5 &=& {\cal C}_{15} = - \frac{1}{2} 
    \left(\frac{\partial p_1}{\partial d_1} \right)_{\mbox{i}}
        > 0, 
        \label{eqn:C_5}
\end{eqnarray}
and ${\cal C}_{14}$ (\ref{eqn:C_n4}) does not exist because 
$\delta A_1=0$ (\ref{eqn:delta_A_n}) and $\delta \sigma_1=0$
((\ref{eqn:delta_F^i_N0}) and (\ref{eqn:delta_sigma_n})). 
Although equation (\ref{eqn:delta_F_2}) does not
dictate the sign of $\delta_2$,
we only consider the case of $\delta_2 \leq 0$ 
because of the two-step deformation
(Figs. \ref{fig:min_dist} and \ref{fig:bending_myelin}).
The terms ${\cal B}$ and ${\cal C}_0$ 
originate from the curvature energy $F^{\mbox{\scriptsize c}}$ (\ref{eqn:F^c_0}),
${\cal C}_1$ from the elastic energy $F^{\mbox{\scriptsize e}}$ (\ref{eqn:F^e}), 
and ${\cal C}_2$, ${\cal C}_3$, and ${\cal C}_5$ from the inter-bilayer 
interaction energy $F^{\mbox{\scriptsize i}}$ (\ref{eqn:F^i}).
Here we assume that the outer bilayer is under tension 
($\bar{\sigma}_2 >0$) and  the inter-bilayer interaction is 
repulsive ($\bar{p}_1 >0$).  
We expect ${\cal C}_3 < 0$ and ${\cal C}_5 > 0$ because  the 
inter-bilayer repulsion increases as the bilayer 
tension $\sigma_2$ or the inter-bilayer spacing $d_1$
decreases, i.e., 
$\partial f^{\mbox{\scriptsize i}}_1/\partial \sigma_2 < 0$
and $(\partial p_1/\partial d_1) < 0$.
Therefore, ${\cal C}_1$ and ${\cal C}_2$
 contribute to coiling, while 
 ${\cal C}_3$ and ${\cal C}_5$ hinder coiling.

%
\subsection{Criterion of Coiling Instability
            \label{sec:criterion}
            }
Equation (\ref{eqn:delta_F_2}) gives rise to a simple
stability criterion 
when all the ${\cal O}(\delta_2^2)$ terms are negligible:
 a straight, two-bilayer myelin is unstable 
against bending or coiling if there exists a set of
$(\delta_2, R_1)$ such that 
\begin{equation}
\delta F \sim {\cal B} + \frac{\delta_2}{R_1}
    \left[{\cal C}_1+{\cal C}_2+{\cal C}_3\right]  < 0,
    \label{eqn:criterion_2}
\end{equation}
where ${\cal C}_0$ is dropped because
 ${\cal B} \gg |{\cal C}_0\delta_2 /R_1 |$ 
 (see the text below (\ref{eqn:bar_tilde_kappa_n})).
  By setting $\delta F=0$ we can define a threshold 
 of $\delta_2$, denoted as $\delta_{\mbox{\scriptsize th}}$,
 for coiling instability.  If one can find a finite $R_1$ 
 and a negative $\delta_2$ of
 magnitude larger than $|\delta_{\mbox{\scriptsize th}}|$,
then state (iii) is energetically more favorable than 
state (i) (Fig. \ref{fig:bending_myelin}).
However, $\delta_2$ and $R_1$ must be bounded by the
geometry of the myelin:  for example, $|\delta_2|$ must be 
less than $(\bar{r}_2-\bar{r}_1)$, 
and $R_1$ must be greater than $r_2$.
The equilibrium value of $(\delta_2, R_1)$ can  
be determined when the terms of ${\cal O}(\delta_2^2)$ in 
  (\ref{eqn:delta_F_2}) are known.

If the  tension $\bar{\sigma}_2$ is large 
enough so that ${\cal C}_1 \gg |{\cal C}_2+{\cal C}_3|$,
i.e., the inter-bilayer interaction is negligible,
 criterion (\ref{eqn:criterion_2}) is  reduced to
\begin{equation}
    \delta F \sim {\cal B} + \frac{\delta_2}{R_1} {\cal C}_1 < 0.
   \label{eqn:criterion_2_s}
\end{equation}
Furthermore, 
when $\bar{r}_2\approx \bar{r}_1\gg (\bar{r}_2- \bar{r}_1)$, 
 we expect ${\cal B}\approx \kappa/R_1^2$ 
(see (\ref{eqn:bar_tilde_kappa_n}) and (\ref{eqn:B_n})) and
${\cal C}_1\approx \bar{\sigma}_2$.
In such case the  $\delta_2$ threshold for coiling 
 is given by 
\begin{equation}
 \delta_{\mbox{\scriptsize th}}= -\frac{\kappa}{\bar{\sigma}_2 R_1}.
    \label{eqn:delta_th}
\end{equation}
Now we investigate a two-bilayer myelin composed of 
typical lipid bilayers with parameters 
\begin{eqnarray}
    \bar{r}_2 &=& 20\; \mu\mbox{m (Fig. \ref{fig:myelin_structure})},
    \nonumber \\
    \bar{d}_1 &=& \bar{r}_2 - \bar{r}_1 = 2.4\;\mbox{nm \cite{Rand-1989}},
    \nonumber \\
    R_1 &=& 30\; \mu\mbox{m \cite{Haran-2002,Sakurai-1985,Frette-1999}},
    \nonumber \\
    \kappa &=& 10^{-12}\; \mbox{erg \cite{Helfrich-1984}},
    \nonumber \\
    dp^{\mbox{\scriptsize w}}_2 &=& 0 \;\mbox{dyne/cm$^2$ (Eq. (\ref{eqn:force_n}))},
    \nonumber \\
    \bar{p}_1 & =& 10^3\;\mbox{dyne/cm$^2$ (Eq. (\ref{eqn:bar_p_n})) \cite{Zou-2006,Huang-2005}},
    \nonumber 
\end{eqnarray}
and the temperature $T = 4.1 \times 10^{-14}$ erg (300 K).
For simplicity the water pressure difference 
$dp^{\mbox{\scriptsize w}}_2$ is assumed to be zero. 
The values of these parameters are carefully 
chosen such that the results of this example can be readily
generalized for myelins of many bilayers 
 (Sect. \ref{sec:generalization}).
Equation (\ref{eqn:force_n}) with the above parameters implies 
${\cal C}_1 \approx \bar{\sigma}_2\approx \bar{r}_2 \bar{p}_1 = 2$ dyne/cm 
\cite{Evans-1987,Evans-1990} and thus 
$\delta_{\mbox{\scriptsize th}}\approx -0.02 \mbox{ \AA}$
according to (\ref{eqn:delta_th}).
Because $|\delta_{\mbox{\scriptsize th}}|$ is much less than
typical bilayer thickness or inter-bilayer spacing 
(Fig. \ref{fig:myelin_structure}) \cite{Rand-1989},
 there exists a negative $\delta_2$ such that 
$|\delta_2| > |\delta_{\mbox{\scriptsize th}}|$.
Therefore state (iii) is more stable than state (i)
 (see (\ref{eqn:criterion_2}) and 
 Fig. \ref{fig:bending_myelin}).  This numerical example 
 suggests that criterion (\ref{eqn:criterion_2_s})
can be met easily and  a two-bilayer myelin 
may bend or coil if the lateral tension of the outer 
bilayer is sufficiently large. 
However, two questions arise with regard to the above example.
Can we ignore the ${\cal O}(\delta_2^2)$ terms in 
equation (\ref{eqn:delta_F_2})? Moreover, 
is the inter-bilayer interaction negligible, i.e.,
is it possible that ${\cal C}_1 \gg |{\cal C}_2+{\cal C}_3|$?
Because the complete functional form of 
$f_n^{\mbox{\scriptsize i}}$ (\ref{eqn:f_n^i})
is not known, it seems difficult, if not impossible, 
to answer these two questions with logical rigor.
In the next subsection we will try to address these issues 
by means of some  physical arguments.

In fact, equation (\ref{eqn:delta_th}) can also be viewed 
from another perspective: it provides a lower bound for 
$\bar{\sigma}_2$ in a coiled myelin
when $|\delta_{\mbox{\scriptsize th}}|$ is replaced 
with the inter-bilayer spacing, i.e., 
a coiled two-bilayer myelin with coiling curvature $1/R_1$ 
(\ref{eqn:R_n}) must have a $\bar{\sigma}_2$ greater than 
$\kappa/(\bar{d}_1 R_1)$.  The actual  $\bar{\sigma}_2$ 
 is expected to be larger than this value because of 
 the contributions from the inter-bilayer interaction,
the higher-order terms in $E^{\mbox{\scriptsize c}}_n$ (\ref{eqn:E^c_n}), and 
the ${\cal O}(\delta_2^2)$ terms in (\ref{eqn:delta_F_2}).
In principle, we can obtain a more sophisticated lower
bound for  $\bar{\sigma}_2$ using equation 
(\ref{eqn:delta_F_2}) if the functional form of $f^{\mbox{\scriptsize i}}_1$ is
known. 

%
\subsection{Validity of the Criterion
    \label{sec:validity}
    }
Equation (\ref{eqn:delta_th}) is valid  only when 
(a) $\bar{r}_2\approx \bar{r}_1\gg (\bar{r}_2- \bar{r}_1)$,
(b) ${\cal C}_1 \gg |{\cal C}_2 + {\cal C}_3|$, i.e.,
the inter-bilayer interaction is negligible,
and
(c) the ${\cal O}(\delta_2^2)$ terms of 
(\ref{eqn:delta_F_2}) can be ignored.
Condition (a) is obviously met by the numerical 
example given below (\ref{eqn:delta_th}).
Intuitively, condition (b) would be satisfied if
the bilayer tension $\bar{\sigma}_2$ is sufficiently large.
For a straight two-bilayer myelin,  
$\bar{\sigma}_2$ is balanced by the net pressure
difference $\Delta p$
across the outer bilayer cylinder (\ref{eqn:force_n}):
$\bar{\sigma}_2=\bar{r}_2 \Delta p 
\approx \bar{r}_2 ( dp^{\mbox{\scriptsize w}}_2 + \bar{p}_1 )$,
where  $\bar{p}_1$ (\ref{eqn:bar_p_n}) 
represents the strength of the inter-bilayer interaction. 
This force balance equation implies that 
$\bar{\sigma}_2$ increases with $\bar{r}_2$ or $\Delta p$. 
Because $\bar{p}_1$ ($>0$) tends to decrease 
as $\bar{\sigma}_2$ increases,       
we expect  a sufficiently large $\bar{r}_2$ or 
$dp^{\mbox{\scriptsize w}}_2$ 
 would yield a $\bar{\sigma}_2$ large enough so that 
 the inter-bilayer interaction can be neglected.
In the following we will examine closely whether  
conditions (b) and (c) are satisfied by the 
example below (\ref{eqn:delta_th}).
Our strategy is to  compare 
the orders of magnitude of the terms of interest.

Equations (\ref{eqn:delta_d_n}) and (\ref{eqn:delta_A_n}) with
$\delta_{\mbox{\scriptsize th}}\approx -0.02 \mbox{\AA}$ 
yield
\begin{equation}
    \frac{|\delta_{\mbox{\scriptsize th}}|}{R_1} \sim 
    \frac{\delta d_1}{R_1} \sim
    \frac{\delta A_2}{\bar{A}_2} \sim 10^{-7}.
    \label{eqn:A_2_ratio}
\end{equation}
Here the use of symbol ``$\sim$'' means we are only 
interested in the orders of magnitude of the terms
on its both sides.  
The above equation in turn implies that the 
${\cal O}(\delta_2^2)$ terms in $\delta d_1$ 
(\ref{eqn:delta_d_n}) are negligible.
Experimental results suggest that the bilayer tension varies
linearly with the fractional change in bilayer area 
when it is sufficiently large ($\geq 0.5$ dyne/cm
for typical lipid bilayers) \cite{Evans-1987,Evans-1990}. 
Thus, in our case
\begin{equation}
  \left(\frac{\partial \sigma_2}{\partial A_2} \right)_{\mbox{i}}
   \bar{A}_2 \approx K_{\mbox{\scriptsize A}},
    \label{eqn:sa}
\end{equation}
where $K_{\mbox{\scriptsize A}}$ is the area compressibility 
modulus of the bilayer. For typical lipid bilayer
membranes,  $K_{\mbox{\scriptsize A}}\approx 200$ dyne/cm 
\cite{Evans-1987}.
Equations (\ref{eqn:delta_sigma_n}) and (\ref{eqn:sa})
imply that the fractional change in $\sigma_2$ is also small:
\begin{equation}
 \frac{\delta\sigma_2}{\bar{\sigma}_2} \approx 
 \frac{K_{\mbox{\scriptsize A}}}{\bar{\sigma}_2}
 \frac{\delta A_2}{\bar{A}_2}\sim 10^{-5},
     \label{eqn:sigma_2_ratio}
\end{equation}
where $\bar{\sigma}_2 = 2$ dyne/cm.
By virtue of  (\ref{eqn:A_2_ratio}) and (\ref{eqn:sa}),
it is straightforward to show that 
in $\delta F^{\mbox{\scriptsize e}}$ (\ref{eqn:delta_F^e_N}), 
the ${\cal O}(\delta_2^2/R_1^2)$   term
is much less than the ${\cal O}(\delta_2/R_1)$  term 
and thus can be ignored.

The change in the inter-bilayer interaction potential 
$f^{\mbox{\scriptsize i}}_1$ reads 
(see (\ref{eqn:f_n^i}) and (\ref{eqn:delta_F^i_N0}))
\begin{equation}
\delta f^{\mbox{\scriptsize i}}_1 = \sum_{k=2}^{6}  {\cal D}_k 
        + {\cal O}\left(\frac{\delta_2^3}{R_1^3}, \delta_2^6\right),
    \label{eqn:delta_f^i}
\end{equation}
where, with (\ref{eqn:p_n}) and (\ref{eqn:bar_p_n}),
\begin{eqnarray}
    {\cal D}_2 &=& -\bar{p}_1 \delta d_1,
        \nonumber \\
    {\cal D}_3 &=& \left(\frac{\partial f^{\mbox{\scriptsize i}}_1}{\partial \sigma_2} \right)_{\mbox{i}}\delta \sigma_2,
        \nonumber \\
    {\cal D}_4 &=& \left(\frac{\partial^2 f^{\mbox{\scriptsize i}}_1}{\partial {\sigma_2}^2} \right)_{\mbox{i}}\frac{\delta \sigma_2^2}{2},
        \nonumber \\
   {\cal D}_5 &=& -\left(\frac{\partial p_1}{\partial d_1} \right)_{\mbox{i}}\frac{\delta d_1^2}{2},
        \nonumber \\
    {\cal D}_6 &=& -\left(\frac{\partial p_1}{\partial {\sigma_2}} \right)_{\mbox{i}}\delta d_1 \delta \sigma_2,
        \nonumber
\end{eqnarray}
and $\delta d_1  \sim |\delta_{\mbox{\scriptsize th}}| \approx 0.02\, \mbox{\AA}$
in our case.
The first-order terms ${\cal D}_2$ and 
${\cal D}_3$ lead to ${\cal C}_2$ (\ref{eqn:C_2}) and 
${\cal C}_3$ (\ref{eqn:C_3}), respectively 
(see (\ref{eqn:delta_F^i_N0})).
The second-order terms ${\cal D}_4$, ${\cal D}_5$,  
and ${\cal D}_6$  contribute to the 
${\cal O}(\delta_2^2)$ terms in (\ref{eqn:delta_F_2}).
In our case
${\cal C}_1\approx 2$ erg/cm$^2$ (\ref{eqn:C_1}) 
is certainly much larger than 
${\cal C}_2 = 2.4 \times 10^{-4}$ erg/cm$^2$  (\ref{eqn:C_2})
because $\bar{r}_2 \gg (\bar{r}_2-\bar{r}_1)$ and 
$\bar{\sigma}_2 \approx \bar{r}_2\bar{p}_1$ 
(\ref{eqn:force_n}).
Therefore ${\cal C}_2$ and ${\cal D}_2$ can be neglected.
Besides, we expect that 
\begin{equation}
    f^{\mbox{\scriptsize i}}_1 \sim p_1 \lambda
    \;\;\;\,\mbox{and}\;\,
    \left(\frac{\partial p_1}{\partial d_1} \right)_{\mbox{i}} \sim
    \frac{p_1}{\lambda},
    \label{eqn:f^i_form}
\end{equation}
where the length $\lambda$ can be as small as the decay 
length ($2\mbox{ -- }3 \, \mbox{\AA}$) of hydration pressure 
\cite{Rand-1989} or as large as the inter-bilayer spacing 
(several nanometers) if the Helfrich repulsion 
\cite{Helfrich-1978} dominates.
This means in our case
\begin{equation}
    \frac{\delta  d_1}{\lambda} \sim 
    \frac{|\delta_{\mbox{\scriptsize th}}|}{\lambda} 
    \sim 10^{-3}\mbox{ -- }  10^{-2} \ll 1 
\label{eqn:delta_d_lambda}
\end{equation}
and $|\delta_{\mbox{\scriptsize th}}| {\cal C}_1/R_1\sim 10^{-7}$ 
dyne/cm is much larger than 
${\cal D}_5\sim {\cal C}_5 \delta_{\mbox{\scriptsize th}}^2 \sim 10^{-9}$ 
dyne/cm (see (\ref{eqn:delta_F_2})).
Hence we can ignore 
 those ${\cal O}(\delta_2^2)$ terms of (\ref{eqn:delta_F_2})
 arising from ${\cal D}_5$ .
The functional dependence of $f^{\mbox{\scriptsize i}}_1$
on $\sigma_2$ is necessary in order to decide whether 
${\cal D}_3$, ${\cal D}_4$, and ${\cal D}_6$ are negligible. 
To go further, we consider two extreme situations.
The actual $f^{\mbox{\scriptsize i}}_1(\sigma_2)$  
is expected to fall in between these two extremes.

In the first extreme situation, we assume ${\sigma}_2$ 
is large enough so that the bilayer thermal undulations 
\cite{Helfrich-1984} can be ignored. 
This assumption implies that 
${\sigma}_2$ is coupled with $f^{\mbox{\scriptsize i}}_1$  
mainly via surfactant number density $\rho_{\mbox{\scriptsize s}}$
(number of surfactant molecules per unit bilayer area)
of the outer bilayer.
Intuitively, when $\sigma_2$ is sufficiently large
 (close to the lysis tension) \cite{Evans-1987}, 
 $f^{\mbox{\scriptsize i}}_1$ should be
 roughly proportional to $\rho_{\mbox{\scriptsize s}}$.
With fixed inter-bilayer spacing $d_1$
(see (\ref{eqn:delta_f^i})), 
equations (\ref{eqn:sigma_2_ratio}), (\ref{eqn:f^i_form}),
and the assumption of
$f^{\mbox{\scriptsize i}}_1 \propto \rho_{\mbox{\scriptsize s}}$  
lead to
\begin{equation}
    \frac{\delta p_1}{p_1} \sim
    \frac{\delta f^{\mbox{\scriptsize i}}_1}{f^{\mbox{\scriptsize i}}_1}
    \sim - \frac{\delta A_2}{A_2}    
    \approx - \frac{\sigma_2}{K_{\mbox{\scriptsize A}}} \frac{\delta\sigma_2}{{\sigma}_2}
\label{eqn:f_p_ratio}
\end{equation}
and thus
\begin{equation}
\frac{\partial f^{\mbox{\scriptsize i}}_1}{\partial \sigma_2}
\approx \frac{\delta f^{\mbox{\scriptsize i}}_1}{\delta \sigma_2}
= - Y \frac{f^{\mbox{\scriptsize i}}_1}{\sigma_2},
    \label{eqn:ratio_f_sigma}
\end{equation}
where 
\begin{equation}
 Y = - \frac{\delta f^{\mbox{\scriptsize i}}_1/f^{\mbox{\scriptsize i}}_1}{\delta \sigma_2/{\sigma}_2}
\approx - \frac{\delta f^{\mbox{\scriptsize i}}_1/f^{\mbox{\scriptsize i}}_1}{\delta A_2/A_2}
\frac{\sigma_2}{K_{\mbox{\scriptsize A}}}
    \sim \frac{\sigma_2}{K_{\mbox{\scriptsize A}}}.
    \label{eqn:Y}
\end{equation}
In our case $\sigma_2 = 2$ dyne/cm,
$K_{\mbox{\scriptsize A}} \approx 200$ dyne/cm,
and hence $Y \sim 10^{-2}$.
By use of (\ref{eqn:f^i_form}), (\ref{eqn:f_p_ratio}), and 
(\ref{eqn:ratio_f_sigma}), we can rewrite 
${\cal D}_3$, ${\cal D}_4$, and ${\cal D}_6$:
\begin{eqnarray}
{\cal D}_3 &= & \delta  f^{\mbox{\scriptsize i}}_1
         \approx - Y f^{\mbox{\scriptsize i}}_1
    \left( \frac{\delta \sigma_2}{\bar{\sigma}_2} \right)
    \sim Y \bar{p}_1 \lambda
    \left( \frac{\delta \sigma_2}{\bar{\sigma}_2} \right). 
        \nonumber \\
{\cal D}_4 &=& \delta \left(\frac{\partial f^{\mbox{\scriptsize i}}_1}{\partial \sigma_2} \right)\frac{\delta \sigma_2}{2}
        \sim Y \;\delta \left(\frac{f^{\mbox{\scriptsize i}}_1}{\sigma_2}\right)\delta \sigma_2
    \nonumber \\
    &\sim & Y (Y-1) \bar{p}_1 \lambda 
   \left(\frac{\delta \sigma_2}{\bar{\sigma}_2}\right)^2.
     \nonumber \\
{\cal D}_6 &=& - \delta p_1 \delta d_1 
    \sim \left(\frac{\delta p_1}{p_1}\right) 
    \left( \frac{\delta d_1}{\lambda}\right) \bar{p}_1 \lambda
    \nonumber \\
    & \sim & Y \bar{p}_1 \lambda 
    \left(\frac{\delta \sigma_2}{\bar{\sigma}_2}\right)
     \left( \frac{\delta d_1}{\lambda}\right).
    \nonumber 
\end{eqnarray}
Given $Y \sim 10^{-2}$, $\bar{\sigma}_2 \approx 2$ dyne/cm, 
$K_{\mbox{\scriptsize A}} \approx 200$ dyne/cm,  
$\bar{p}_1 = 10^3$ dyne/cm$^2$, and $\lambda \sim 1$ nm, 
equations (\ref{eqn:delta_F^e_N}), (\ref{eqn:sigma_2_ratio}), 
and (\ref{eqn:delta_d_lambda}) imply
\[
 \frac{|\delta F^{\mbox{\scriptsize e}}|}{A_2} 
 = \frac{\bar{\sigma}_2 |\delta A_2|}{A_2}
 \approx \frac{\bar{\sigma}_2^2}{K_{\mbox{\scriptsize A}}}
         \frac{|\delta \sigma_2|}{\bar{\sigma}_2}
 \gg |{\cal D}_3|
 \gg |{\cal D}_4| \;\,\mbox{or}\;\, |{\cal D}_6|.
\]
Therefore,  ${\cal C}_3$ 
(\ref{eqn:C_3}), which arises from ${\cal D}_3$, 
as well as those ${\cal O}(\delta_2^2)$ terms  of
(\ref{eqn:delta_F_2}) contributed by
${\cal D}_4$ and ${\cal D}_6$ are negligible
in the example  below (\ref{eqn:delta_th}).
We also note that with (\ref{eqn:C_3}), (\ref{eqn:sa}), 
(\ref{eqn:f^i_form}), (\ref{eqn:ratio_f_sigma}),
and (\ref{eqn:Y}), 
\begin{eqnarray}
{\cal C}_3 & \approx & 
\left(\frac{\partial f_1^{\mbox{\scriptsize i}}}{\partial \sigma_2} \right)_{\mbox{i}}
    K_{\mbox{\scriptsize A}}
    \approx -Y \frac{\bar{f}^i_1}{\bar{\sigma}_2}  K_{\mbox{\scriptsize A}}
    \sim -Y  K_{\mbox{\scriptsize A}} \frac{\bar{p}_1\lambda}{\bar{\sigma}_2} 
    \nonumber \\
 & \sim & \, -\bar{p}_1\lambda \, \sim\,  10^{-5}\mbox{ -- } 10^{-4}
     \mbox { erg/cm$^2$},
     \nonumber
\end{eqnarray}
which has approximately the same order of magnitude as 
${\cal C}_2 = 2.4 \times 10^{-4}$ erg/cm$^2$  (\ref{eqn:C_2}).
This means in our case coiling might even lead to $({\cal C}_2 + {\cal C}_3) >0$, 
i.e., a  decrease in the inter-bilayer interaction $F^i$
(see (\ref{eqn:delta_F^i_N}), (\ref{eqn:delta_F_2}), 
(\ref{eqn:criterion_2}),
and the text below (\ref{eqn:C_5})).

Now we consider the other extreme situation:
 the dependence of  $f^{\mbox{\scriptsize i}}_1$ on
 $\bar{\sigma}_2$  originates from bilayer thermal
undulations \cite{Helfrich-1984,Seifert-1995,Netz-1995}.
Here we adopt the functional form derived by Seifert 
(Eq. (9) of \cite{Seifert-1995}) and express
$f^{\mbox{\scriptsize i}}_1$  as
\begin{equation}
 f^{\mbox{\scriptsize i}}_1 = \hat{f}^{\mbox{\scriptsize i}}_1(\bar{d}_1) +
 \frac{6 b^2 T^2}{\kappa \bar{d}_1^2}\frac{y^2}{\sinh^2(y)},
    \label{eqn:f^i_1}
\end{equation}
where $b=0.1$, $\bar{d}_1=(\bar{r}_2- \bar{r}_1)$, and
$y=(\bar{\sigma}_2/bT)^{1/2} \bar{d}_1/2$. 
The term $\hat{f}^{\mbox{\scriptsize i}}_1$ 
is the part of $f^{\mbox{\scriptsize i}}_1 $ that is 
independent of $\bar{\sigma}_2$.  The last term  is just
the Helfrich potential with variable membrane tension. 
The  dependence on $\bar{\sigma}_1$ is neglected because 
$\bar{\sigma}_1$ is unchanged in the two-step deformation.
With (\ref{eqn:f^i_1}) and 
the parameters given below (\ref{eqn:delta_th}),
straightforward calculations show that
${\cal C}_1 \approx 2\; \mbox{erg/cm$^2$}$ (\ref{eqn:C_1})
is much larger than 
$|{\cal C}_3| \approx 0.04\; \mbox{erg/cm$^2$}$ (\ref{eqn:C_3})
and that both ${\cal D}_4$ and ${\cal D}_6$ are  much less 
than ${\cal D}_3$. 
Therefore,  ${\cal C}_3$ as well as 
the ${\cal O}(\delta_2^2)$ terms
of  (\ref{eqn:delta_F_2}) associated with  ${\cal D}_4$ 
and ${\cal D}_6$ are negligible.

In the previous subsection we use equation 
(\ref{eqn:delta_th}) to investigate the coiling 
instability  of  a two-bilayer myelin 
(see (\ref{eqn:criterion_2_s})). 
This equation is valid only when 
(a) $\bar{r}_2\approx \bar{r}_1 \gg (\bar{r}_2 - \bar{r}_1)$, 
(b) ${\cal C}_1\gg |{\cal C}_2+{\cal C}_3|$, 
i.e., the inter-bilayer interaction is negligible, and
(c) the ${\cal O}(\delta_2^2)$  terms in 
(\ref{eqn:delta_F_2}) can be ignored.
We are interested in the example described 
below (\ref{eqn:delta_th}), which clearly 
satisfies condition (a).
In this subsection we examine conditions (b) and (c)
by estimating the orders of magnitude
of the relevant terms.
We show that these two conditions are also satisfied. 
Thus, equation (\ref{eqn:delta_th}) is indeed applicable to 
the  case of interest.

%
\subsection{Generalization for Multiple-bilayer Myelins
            \label{sec:generalization}
            }
In principle the results of Section 
\ref{sec:criterion} can be applied to 
myelins of multiple bilayers.
 This is because in terms of energy 
(see (\ref{eqn:F}) and (\ref{eqn:delta_F_N})), 
an $N$-bilayer myelin can be decomposed into 
$(N-1)$ nested two-bilayer myelins.
Although  the bilayer tension $\sigma_n$
(\ref{eqn:sigma_n}) may vary from bilayer to bilayer,
equation (\ref{eqn:delta_F_N}) implies that
the coiled state is still  favorable 
 as long as the net decrease in $F^{\mbox{\scriptsize e}}$ 
is larger than the net increase 
in $F^{\mbox{\scriptsize c}}+F^{\mbox{\scriptsize i}}$.

Equation (\ref{eqn:delta_F_N}) and its implications for 
myelin coiling can also be understood with a gedanken 
experiment, where a multiple-bilayer myelin is constructed 
from  two-bilayer myelins. 
We take the numerical example below (\ref{eqn:delta_th})
 as our initial two-bilayer myelin.
The bending curvature $1/R_1$ and the eccentricity $\delta_2$
are allowed to vary freely so that the myelin is in 
equilibrium. The equilibrium state has a finite $R_1$
 and  a $\delta_2$ of magnitude larger than
$|\delta_{\mbox{\scriptsize th}}|$ ($\approx 0.02 \mbox{\AA}$).
This means $\delta F$ (\ref{eqn:delta_F_2}) is less
than zero.
Put it in another way, the two-step deformation 
(Fig. \ref{fig:bending_myelin}) 
produces a  surplus of energy.
Since $|\delta_{\mbox{\scriptsize th}}|$ of this initial 
myelin is quite small, the equilibrium value of $|\delta_2|$ 
is expected to be much larger than $|\delta_{\mbox{\scriptsize th}}|$.
Therefore the energy surplus  $|\delta F|$ should also
 be much larger than the curvature energy cost of coiling 
$\delta F^{\mbox{\scriptsize c}}$ (\ref{eqn:delta_F^c_N}).
We keep $R_1$ unchanged and insert another two-bilayer 
myelin into the inside of the initial one. 
Now the system becomes a four-bilayer myelin.
The energy surplus $|\delta F|$ may 
increase or decrease as a result of this new addition.
We can repeat this process and add more 
two-bilayer myelins to the system as long as the energy
surplus is not depleted, i.e., $\delta F$ is still less 
than zero.
Even when $\delta F>0$, we can increase the bilayer tensions
in those existing two-bilayer myelins so that  $\delta F$ 
becomes negative again.
The above gedanken experiment shows clearly that 
a multi-bilayer myelin can coil to lower its energy when
the bilayer tension is sufficiently large.

Let us consider the case depicted by 
Figure \ref{fig:contact}b.
In this case the material influx density 
for the core is increasingly greater than that for the shell. 
Eventually the bilayer tension in the shell will be large
enough to cause the myelin to bend or coil.
To be more specific, the free energy analysis 
(\ref{eqn:delta_F_N})  can be 
done in the following manner. 
First, the eccentricities $\delta_n$ (Fig. \ref{fig:min_dist}) 
of the bilayers in the core are artificially constrained to 
be zero, while those in the shell are free to vary.
We start with state (i) (Fig. \ref{fig:bending_myelin}),
i.e., $R_1\rightarrow\infty$ and $\delta_n=0$ for all $n$.
By symmetry state (i) is  the 
equilibrium configuration for straight myelins.  
Secondly, we perform the two-step deformation 
(Fig. \ref{fig:bending_myelin}) and allow the 
system to reach equilibrium at finite $R_1$.
This means those variable eccentricities $\delta_n$
are at the values minimizing
the total energy $F$ (\ref{eqn:F}).
Because every $\delta_n$ in the core is set to zero, the 
deformation of the core only costs curvature energy 
 (\ref{eqn:delta_F^c_N}).
Previous analysis of the two-bilayer myelin suggests that
when the bilayer tensions $\sigma_n$ in the shell are 
large enough, the two-step deformation would lead to 
a net decrease in $F$ and thus state (iii) becomes 
energetically favorable.
Lastly, the constraint of $\delta_n=0$ in the core
is dismissed and  the system is allowed to equilibrate again.
Obviously, if state (iii) has been more stable than 
state (i) as a result of large bilayer tension in the shell,
it would be still favorable
because the new degrees of freedom would only lower 
the total energy $F$ further.
Therefore, we conclude that the myelin of 
Figure \ref{fig:contact}b would bend or coil when
the bilayer tension in the shell is sufficiently large.
In the sense of coarse-graining, 
this case can be treated as a ``two-bilayer'' myelin, 
whose inner and outer bilayers
correspond to the core  and the shell, respectively.

%
%
%
\section{Discussion \label{sec:Discussion}}
Our model predicts that a sufficiently 
large bilayer tension may cause a myelin to coil. 
The coiling mechanism is explained 
with the two-bilayer case studied in detail 
in the previous section.
In the following we will discuss the  experimental 
techniques that might be used to test our model.
We will also look into some important features
as well as implications of our model.

In order to test the proposed model of myelin coiling,
one needs to be able
to either measure or manipulate the bilayer tension of
a myelin.  Optical tweezers may be a promising tool
\cite{Bar-Ziv-1994,Curtis-2002}.
When a bilayer is optically tweezered, surfactant molecules 
are pulled into the trap set by the tweezers and thus the
bilayer tension increases.
An advantage of optical tweezers is that they can 
penetrate into a myelin and grab multiple bilayers.
However, laser-induced tension is usually weak
 ($10^{-3}-10^{-2}$ dyne/cm) \cite{Bar-Ziv-1998}. 
 It  remains a question
 whether such a weak tension can induce myelin coiling.
 Another method  is micropipet suction 
 \cite{Evans-1987,Evans-1990}. 
 In contrast to optical tweezers, this method can produce 
 a bilayer tension as large as the lysis 
 tension \cite{Evans-1987}.
Nonetheless, it can be applied only from the outside
 of a myelin. 
 Although  suction of a myelin with a micropipet 
 is an interesting problem in its own right, 
 whether it can cause myelin coiling is still uncertain.

The second constraint described in Section \ref{sec:modeling}
requires volume conservation for each bilayer cylinder.
It should be a fair condition for defect-free myelins 
 \cite{Kleman-1983}
since  water permeabilities of typical lipid bilayers 
are quite small \cite{Huang-2005,Marrink-1994}.
This constraint is also necessary in our model
because it  eliminates a potential 
inconsistency between properties (a) and (b) described in
Section \ref{sec:modeling}: 
if bilayers can exchange material, a nonuniform bilayer 
tension may induce a sufficiently large material flow 
between bilayers to relax the tension that drives coiling.

For simplicity the spontaneous curvature $K_0$  
(\ref{eqn:E^c}) is set to zero in this work.  
A nonzero or nonuniform $K_0$ would only alter 
the curvature energy $F^{\mbox{\scriptsize c}}$ (\ref{eqn:F^c_0}) 
by some finite amount and therefore does not weaken 
the proposed coiling mechanism at all 
\cite{Frette-1999,Santangelo-2002}.
In addition, the cross sections of the bilayer cylinders 
 are assumed to be circular (Fig. \ref{fig:min_dist}).
Although this might be a fair assumption for straight 
myelins, there is no reason why this is still the case
for coiled myelins. Because $\delta_n$ is in general nonzero
in state (iii) (Fig. \ref{fig:bending_myelin}), 
the shape of the cross section should change accordingly.
However, allowing this to happen would only lower the energy 
of a coiled myelin further (cf. (\ref{eqn:delta_F_N})) and
thus would not render our model invalid.

The myelin free energy $F$  (\ref{eqn:F}) is the sum of 
the curvature energy $F^{\mbox{\scriptsize c}}$, the elastic energy $F^{\mbox{\scriptsize e}}$ 
of bilayer tension, and the inter-bilayer
interaction energy $F^{\mbox{\scriptsize i}}$.
Because  $F^{\mbox{\scriptsize c}}$ is known (\ref{eqn:F^c_0}), 
determining the change in $F^{\mbox{\scriptsize c}}$ due 
to coiling is straightforward (\ref{eqn:delta_F^c_N}).
The change in $F^{\mbox{\scriptsize e}}$, to lowest order, is given by 
$\sum \bar{\sigma}_n \delta A_n$, 
where $\bar{\sigma}_n$ is taken as an independent 
parameter (see (\ref{eqn:delta_F^e_N})).
Thus, direct knowledge of $F^{\mbox{\scriptsize e}}$ is not crucial 
for studying myelin coiling at the level of this work.
This, however, is not the case for $F^{\mbox{\scriptsize i}}$ (\ref{eqn:F^i}).
In order to analyze the coiling instability in detail,
the  functional form of $f^{\mbox{\scriptsize i}}_n$ (\ref{eqn:f_n^i}) is 
required  even though we have shown that its
contributions can be neglected under certain 
conditions (Sect. \ref{sec:coiling_2_myelin}).
Yet, to our knowledge, the complete functional 
expression of $f^{\mbox{\scriptsize i}}_n$ is still not known.
Therefore more investigations of $f^{\mbox{\scriptsize i}}_n$ are needed.
In the context of this work, 
$f^{\mbox{\scriptsize i}}_n$  has to satisfy the force balance 
equation (\ref{eqn:force_n}).

In the example below (\ref{eqn:delta_th})
we use a small inter-bilayer pressure 
($\bar{p}_1 = 10^3$ dyne/cm$^2$) 
while our previous work \cite{Huang-2005} 
suggests that the inter-bilayer pressure (\ref{eqn:p_n}) 
in a multi-bilayer myelin is, 
on the average, larger than this value.
This, however, is not a contradiction.
In view of equation (\ref{eqn:force_n}), 
 a two-bilayer myelin with  $\sigma_2=2$ dyne/cm, 
 $r_2=20$ $\mu$m, and $dp^{\mbox{\scriptsize w}}_n =0$ 
 dyne/cm$^2$ must have an inter-bilayer pressure 
 of about $10^3$ dyne/cm$^2$.
As for myelins composed of multiple bilayers,
equation (\ref{eqn:force_n}) also allows a much larger 
inter-bilayer pressure $p_n$, provided that the net pressure 
difference $(dp^{\mbox{\scriptsize w}}_n + p_{n-1} - p_n)$ 
yields a reasonable bilayer tension $\sigma_n$.
Thus, a multi-bilayer myelin may have a large 
inter-bilayer pressure $p_n$ as long as the net pressure 
gradient across the bilayers is small.
This leads to an interesting question in the physics of 
myelins: what determines the width (diameter) of myelins?
Our model suggests  bilayer  tension may be an
important factor.
As just mentioned,
 an inter-bilayer pressure of
$10^3$ dyne/cm$^2$ ($\approx 10^{-3}$ atm) with a radius of
20 $\mu$m yields a bilayer tension of 2 dyne/cm, which is
close to the lysis tension of  lipid bilayer 
membranes ($1\mbox{ -- }20$ dyne/cm) \cite{Evans-1987}. 
In  contact experiments myelins of diameter 
$10\mbox{ -- }50$ $\mu$m
form immediately after bulk surfactant is contacted with 
water (Sect. \ref{sec:contact})
\cite{Sakurai-1990,Buchanan-2000,Haran-2002}.
The surfactant concentration at the contact interface 
should be sufficiently high so that the inter-bilayer 
interaction  is repulsive.
Therefore, even in view of the complex kinetics of 
surfactant dissolution \cite{Warren-2001}, 
an obvious upper bound for myelin diameter is prescribed by 
the lysis tension of the bilayer
and the outermost inter-bilayer pressure in myelins.
By the same token,  we can estimate
the outermost inter-bilayer pressure 
using the observed myelin diameter and lysis tension.

%
%
\section{Conclusions \label{sec:conclusions}}
In this work we propose a model to explain coiling
of myelins often seen in contact experiments.
This model has two important features: the constituent
bilayer cylinders of a myelin can be non-coaxial and
the bilayer  tension can vary from bilayer to 
bilayer. 
The analysis indicates the importance of bilayer tension
in determining  the morphology of a myelin:
 a myelin can  coil if the bilayer  tension is large enough.
The coiling mechanism is in a sense similar to the classical
Euler buckling of a thin elastic rod subject to axial 
compression \cite{Love-1944}.

Our model suggests that myelins have complex internal 
structures. The internal stresses such as bilayer lateral 
tension may be very different for seemingly like myelins. 
As a result, our model should not be deemed inconsistent with 
other models \cite{Mishima-1992,Frette-1999,Santangelo-2002}.
On the contrary, the existence of all these models, 
including ours, reflects the complexity of myelinic 
structures and suggests multiple routes to myelin coiling.
However, the following questions are also legitimate.
Is our model applicable to the systems that were 
previously explained with different models 
\cite{Lin-1982,Mishima-1992,Frette-1999}?
If so, how can we determine which model better explains
the experimental results? 
Moreover, owing to the complexity of myelins, 
experimentalists are faced with a serious challenge: 
how can one produce myelins in a controllable fashion? 
To answer these questions, more experimental as well 
as theoretical investigations are needed.

This work is done as part of the author's doctoral research
  under the supervision of Prof. Thomas A. Witten.
The author thanks Prof. Witten for his generous support 
and countless inspiring discussions.
The author also appreciates many useful comments 
from Ling-Nan Zou and Prof. Sidney R. Nagel.
This work was supported in part by the MRSEC Program 
of the National Science Foundation under Award 
Number DMR-0213745.
%

%
%
%

%
%
%
%
\end{document}